\documentclass[fleqn,usenatbib]{mnras}

\usepackage{newtxtext,newtxmath}
\usepackage[T1]{fontenc}
\usepackage{ae,aecompl}

\usepackage{graphicx}	% Including figure files
\usepackage{amsmath}	% Advanced maths commands
\usepackage{amssymb}	% Extra maths symbols

\title{Image simulations for gravitational lensing with SkyLens}

\author[A. A. Plazas, M. Meneghetti, M. Maturi, \& J. Rhodes]
{
A. A. Plazas,$^{1,2}$\thanks{E-mail: plazasmalagon@gmail.com}M. Meneghetti,$^{3}$\thanks{E-mail: massimo.meneghetti@gmail.com} M. Maturi,$^{4}$\& J. Rhodes$^{1,5,6}$ \\
$^{1}$Jet Propulsion Laboratory, California Institute of Technology, 4800 Oak Grove Dr., Pasadena, CA 91109, USA\\
$^{2}$Astronomical Society of the Pacific, 100 N Main St., Suite 15, Edwarsville, IL 62025, USA \\ 
$^{3}$INAF-Osservatorio di Astrofisica e di Scienze dello Spazio di Bologna, Via Gobetti 93/3, 40129 Bologna, Italy \\
$^{4}$Heidelberg University, Institute of Theoretical Astrophysics, Philosophenweg 12, 69120 Heidelberg, Germany \\
$^{5}$California Institute of Technology, 1200 E. California Blvd., Pasadena, CA 91125, USA \\
$^{6}$Institute for the Physics and Mathematics of the Universe, 5-1-5 Kashiwanoha, Kashiwa, Chiba Prefecture 277-8583, Japan\\}

\pubyear{2018}
\newcommand{\sky}{{\tt{SkyLens}}}
\newcommand{\subf}[2]{%
  {\small\begin{tabular}[t]{@{}c@{}}
  #1\\#2
  \end{tabular}}%
}

\begin{document}
\label{firstpage}
\pagerange{\pageref{firstpage}--\pageref{lastpage}}
\maketitle

\begin{abstract}
We present the latest version of the ray-tracing simulation code \sky, which can be used to develop image simulations that reproduce strong lensing observations by any mass distribution with a high level of realism. Improvements of the code with respect to previous versions include the implementation of the multi-lens plane formalism, the use of denoised source galaxies from the Hubble eXtreme Deep Field, and the simulation of substructures in lensed arcs and images, based on a morphological analysis of bright nearby galaxies. \sky\ can simulate observations with virtually any telescope. We present examples of space- and ground-based observations of a galaxy cluster through the Wide Field Channel on the Advanced Camera for Surveys of the \emph{Hubble Space Telescope}, the Near Infrared Camera of the \emph{James Webb Space Telescope}, the Wide Field Imager of the \emph{Wide Field Infrared Survey Telescope}, the Hyper Suprime Camera of the Subaru telescope, and the Visible Imaging Channel of the \emph{Euclid} space mission. 

%These simulations can be used to validate methods for finding lenses and for modeling their matter distribution.   

\end{abstract}


%%%%%%%%%%%%%%%%%%%%%%%%%%%%%%%%%%%%%%%%%%%%%%%%%%

%%%%%%%%%%%%%%%%% BODY OF PAPER %%%%%%%%%%%%%%%%%%

\section{Introduction}
The phenomenon of gravitational lensing is a direct consequence of the theory of General Relativity, which relates the curvature of spacetime to the distribution of matter and energy in the Universe. The images of distant sources will be magnified and distorted to different degrees depending on the amplitude of the mass concentrations and the relative geometrical configuration between the source, the deflectors encountered by light along its path, and observers on Earth \citep{bartelmann17,treu10,bartelmann01}. As such, gravitational lensing has become a fundamental tool to understand the nature of the dark matter and dark energy that, according to the current standard cosmological model, constitute 95\% of the total content of the Universe \citep{weinberg13,kilbinger15}

The most powerful gravitational lenses are galaxy clusters, since they are the most massive gravitationally bound structures in the universe. Thus, they act as cosmic telescopes, magnifying faint galaxies at high redshifts that contributed to the re-ionization of the early universe \citep{huang16,livermore17,kelly17}. In the central regions of clusters, gravitational lensing can produce multiple images of the same source and elongated arcs. At larger angular distances from the cluster core, the individual tangential distortions induced on images are weak and therefore can only be detected statistically. The combination of these two regimes of gravitational lensing (known as strong and weak lensing, respectively) constrains the dark matter distribution from cluster centers to scales of up to 1 Mpc/h, improving our understanding of the internal structure of galaxy clusters, and as a consequence, of the nature of dark matter and dark energy \citep{allen11}. \textcolor{black}{Upcoming future galaxy surveys surveys will allow the automatic or semi-automatic detection of several gravitational arcs for a systematic study of a large population of cluster-scale lenses \citep{seidel07,stapelberg17,carrasco18,metcalf18}.} 

\textcolor{black}{The measurement of the gravitational lensing signal to the level of precision required by current and future galaxy surveys has proven to be challenging. The characterization of sources of uncertainty such as systematic and modeling errors is crucial. The use of simulations with known inputs and different levels of realism has become a fundamental and standard tool to better understand these errors and validate measurement techniques and algorithms \citep{amara06,heymans06,kitching13,rowe15,li16,birrer18,metcalf18}}.

In this paper we describe new developments and improvements of the gravitational lensing simulation pipeline \sky\ \citep{meneghetti08,meneghetti10a} that can be used to create mock ground- and space-based observations of lensing phenomena by galaxy clusters and by large scale structures in wide fields. Previous versions of \sky\ have been used to study the systematic errors in cluster mass measurements using lensing and X-ray analyses \citep{meneghetti10a,rasia12}. \sky\ has also been used to construct synthetic lenses with the properties of the clusters observed by the \emph{Hubble Space Telescope} (\emph{HST}) in the Frontiers Field Initiative \citep{lotz15} in order to test the accuracy of inversion algorithms that reconstruct the cluster matter distributions from strong lensing measurements \citep{meneghetti17}. {The improvements described in this work are aimed at increasing the realism of the simulations to make \sky\ an even more optimal tool in many applications of lensing by galaxy clusters.}

The paper is organized as follows. In \S\ref{sec:skylens} we review our simulation pipeline and describe in detail the changes with respect to previous versions of the code. As an example of the output of the code, in Section \S\ref{sec:sims} we use the current version of \sky\ to produce simulated observations through several astronomical instruments with realistic parameters and conditions: the Wide Field Channel (WFC) on the Advanced Camera for Surveys (ACS) of \emph{HST}, the Near Infrared Camera (NIRCam) of the \emph{James Webb Space Telescope} (\emph{JWST}), the Wide Field Imager \textcolor{black}{(WFI)} of the \emph{Wide Field Infrared Survey Telescope} (\emph{WFIRST}), the Hyper Suprime Camera (HSC) of the Subaru Telescope, and the Visible Imaging Channel (VIS) of the \emph{Euclid} space mission. We discuss applications of \sky\ in \S\ref{sec:discussion}, and conclude in \S\ref{sec:conclusions}. Unless otherwise noted, we assume a flat $\Lambda$CDM cosmological model with a matter density parameter $\Omega_{\rm{m},0}=0.272$ and a Hubble constant of $H_{0}=70.4$ km/s/Mpc. 

\section{SkyLens}
\label{sec:skylens}
In this section we describe \sky, highlighting the differences with respect to other versions \citep{meneghetti08,meneghetti10a}. The version of \sky\ presented in this paper includes the use of \emph{HST} source galaxy images denoised by the method introduced in \citet{maturi16}. \sky\ now has the capability to include the lensing effects due to multiple lens planes. We also use information from a sample of nearby, well-resolved galaxies to produce realistic simulations of substructures such as regions of active star formation. We model these features as S\'ersic \citep{sersic63} profiles which will be magnified by the lens and appear as knots within the arcs that form at the critical lines in the lens plane. In addition, they can be used by lensing inversion algorithms as additional constraints that must satisfy the condition that they all must trace back to the same source, limiting the parameter space of solutions and increasing the likelihood of a more accurate model optimization in the source plane.

\subsection{General methodology}
\label{sect:genmeth}
For simulating an observation of a patch of the sky, with or without lensing effects, \sky\ goes through the following steps:
\begin{enumerate}
\item it generates a population of galaxies using the luminosity and the redshift distribution of the galaxies in the Hubble Ultra Deep Field (HUDF, \citet{beckwith06});
\item it prepares the virtual observation, receiving the pointing instructions, the exposure time, and the filter, $F(\lambda)$, to be used from the user. The pointing coordinates are used to calculate the level of the background, i.e. the surface brightness of the sky and, in case of observations from the ground, the air mass; 
\item it assembles the virtual telescope. This implies that the user provides a set of input parameters, such as the effective diameter of the telescope, the field of view, the detector specifications (e.g., gain, read-out noise (RON), dark current, and pixel scale) and the additional information necessary to construct the total throughput function, defined as
	\begin{equation}
		T(\lambda)=C(\lambda)M(\lambda)R(\lambda)F(\lambda)A(\lambda) \;.
	\label{eq:transm}
	\end{equation}

In the previous formula, $C(\lambda)$ is the quantum efficiency of the detector, $M(\lambda)$ is the mirror reflectivity, $R(\lambda)$ is the transmission curve of the lenses in the optical system, and $A(\lambda)$ is the extinction function (galactic and atmospheric, in case of simulations of ground based observations);
\item using the spectral energy distributions (SEDs) and the redshifts of the galaxies entering the field of view, it calculates the fluxes in the band of the virtual observation. Then, using the  galaxy templates, the surface brightness of the sources is calculated at each position in the sky and converted into a number of  {\it Analog-to-Digital Units} (ADUs) on the detector pixels.
\item noise is added according to the sky brightness, to the RON, and to the dark current of the detector.  
\end{enumerate}

More precisely, the photon counts on the detector pixels are calculated as the sum of three contributions, namely, from the sky, the galaxies, and the dark current. 
Given a telescope of diameter $D$, the number of photons collected by the detector pixel at $\vec x$ in the
exposure time $t_{\rm exp}$, from a source whose surface brightness is
$I(\vec{x},\lambda)$ (erg s$^{-1}$cm$^{-2}$Hz$^{-1}$arcsec$^{-2}$), is
\begin{equation}
  n_\gamma(\vec x)=\frac{\pi D^2 t_{\rm exp}p^2}{4 h}\int I(\vec
  x,\lambda)\frac{T(\lambda)}{\lambda}{\rm d}\lambda \;,
  \label{eq:ngamma}
\end{equation} 
where $p$ is the pixel size in arcsec, $h$ is the Planck constant, and
$T(\lambda)$ the total transmission given in Eq.~\ref{eq:transm}. The contribution from the sky is given by
\begin{equation}
  n_{\rm sky}=\frac{\pi D^2 t_{\rm exp}p^2}{4 h}\int\frac{T(\lambda)S(\lambda)}{\lambda}{\rm d}\lambda \;,
  \label{eq:nsky}
\end{equation} 
where the $S(\lambda)$ is the sky flux per square arcsec.

The photon counts are converted into ADUs by dividing by the gain $g$:
\begin{eqnarray}
{\rm ADU}_{\rm total}(\vec x)& = & \frac{n_\gamma(\vec x)+n_{\rm sky}+n_{\rm dark}}{g}\nonumber \\
&  = &  {\rm ADU}(\vec x)+{\rm ADU}_{\rm sky} + {\rm ADU_{\rm dark}}\;.
\label{eq:ADUs}
\end{eqnarray}

Photon noise is assumed to be Poisson distributed, with variance 
\begin{eqnarray}
  \sigma_N^2(\vec x)&=&\left\{n_{\rm
  exp}\left(\frac{{\rm RON}}{g}\right)^2+\frac{{\rm ADU_{\rm total}}(\vec x)}{g}\right. \nonumber \\
  & & +\left. \left(f+\frac{a^2}{n_{\rm exp}^2}\right)[{\rm ADU_{\rm total}}(\vec x)]^2\right\} \;.
  	\label{eq:phn}
\end{eqnarray}
In the previous formula, $n_{\rm exp}$
is the number of exposures, and $a$ the flat-field term, which we fix at
$a=0.005$ following \cite{grazian04}. The term $f$ indicates the flat-field
accuracy, which is determined by the number of flat-field exposures and by the
level of the sky background, $B$, as
\begin{equation}
  f=(N_{ff}\cdot B\cdot g)^{-1} \;.
\end{equation}
A more detailed explanation of these formulas is given in Sect. 4.2 of
\cite{grazian04}. 
Thus, in order to simulate the photon noise, we draw random numbers from a Poisson distribution with  variance as given above.

\subsection{Denoised source image generation}

Previous versions of \sky\ have used shapelets \citep{bernstein02,refregier03b} decomposition of HUDF galaxies to simulate a synthetic sky of source galaxies. The current version implements the technique introduced in \citet{maturi16} (based on Expectation Maximization Principal Components Analysis, EMPCA) to obtain a denoised (noise-free) reconstruction of images based on the Hubble eXtreme Deep Field (HXDF) \citep{illingworth13} data, which covers an area of 10.8 arcmin$^2$ down to $\sim$ 30 AB magnitude \citep{rafelski15}.\footnote{The catalog can be found at \url{https://asd.gsfc.nasa.gov/UVUDF/catalogs.html}} These images were taken by the Advanced Camera for Surveys and the Wide-Field Planetary Camera-2 in the {\tt{F435W}}, {\tt{F606W}}, {\tt{F775W}}, {\tt{F814W}}, and {\tt{F850LP}} bands, and have been ``drizzled" to a resolution of 0.03 arc seconds per pixel. The SEDs of the source galaxies were obtained by interpolating the 11 SED templates from the photometric redshift measurements by \citet{coe06}, as determined by the Bayesian Photometric Redshift code \citep{benitez00,benitez04}.\footnote{\url{http://www.stsci.edu/~dcoe/BPZ/}} The \citet{coe06} library includes SEDs for elliptical, spiral (including star-burst), and lenticular galaxies.   

%{\bf Matteo: In reality I used the galaxies located in an effective area of 20 arcmin$^2$ such to avoid the vicinity of the image borders and masks induced by stars and image artefacts.}

%-----

The denoising procedure used to generate the galaxy images used in
this paper implies a linear model
\begin{equation} \label{eqn:model}
  \tilde{g}(\vec{x})=\sum_{k=1}^M a_k \phi_k(\vec{x})
  \;\;\;\mbox{with}\;\;\;
  a_{k}=\sum_i^nd(\vec{x}_i)\phi_{k}(\vec{x}_i) \;,
\end{equation}
describing the postage-stamp image of each galaxy, $d(\vec{x})$, and it is based on a set of orthonormal basis, $\left\{\vec{\phi}_k \;\in\;
\mathbb{R}^{n} \;\mid\; k=1,...,M\right\}$, which is optimally derived
from the data itself. \textcolor{black}{In Eqn.~\ref{eqn:model}, $\tilde{g}(\vec{x})$ is the model of the galaxy we are
interested in, i.e. the denoised image.} The number of components $n$ is equal to the number of pixels in the postage stamps of the galaxies. The basis optimization is based on the Expectation Maximum Principal Component Analysis which captures the
information content in components such that each one of them,
$\Phi_k$, contains more information than the following one,
$\Phi_{k+1}$ \citep{bailey12}. This decomposition allows to select the relevant ones and discard those associated to the noise contribution
\begin{equation}\label{eq:model-split}
  d(\vec{x})=\sum_{k=1}^M a_k \phi_k(\vec{x}) + \sum_{M+1}^{n} a_k \phi_k(\vec{x}) = \tilde{g}(\vec{x}) + \tilde{n}(\vec{x}) \;.
\end{equation}
Here, $\tilde{n}(\vec{x})$ is a
term which contains most of the noise and that we discard. The advantage of the EMPCA with respect to the standard PCA is that it allows to deal with missing data, noise levels varying across the field, and the use of regularization terms. The number of components, $M$, \textcolor{black}{is different for each galaxy, and is chosen through an iterative process that ends when the galaxy model does not change in a significant way. Details on the process and the basis derivation can be found in \citet{maturi16}}.

\begin{figure}
  \centering
  \includegraphics[width=1.0\hsize]{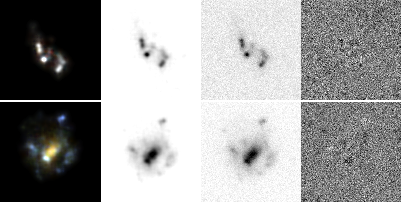}
  \includegraphics[width=1.0\hsize]{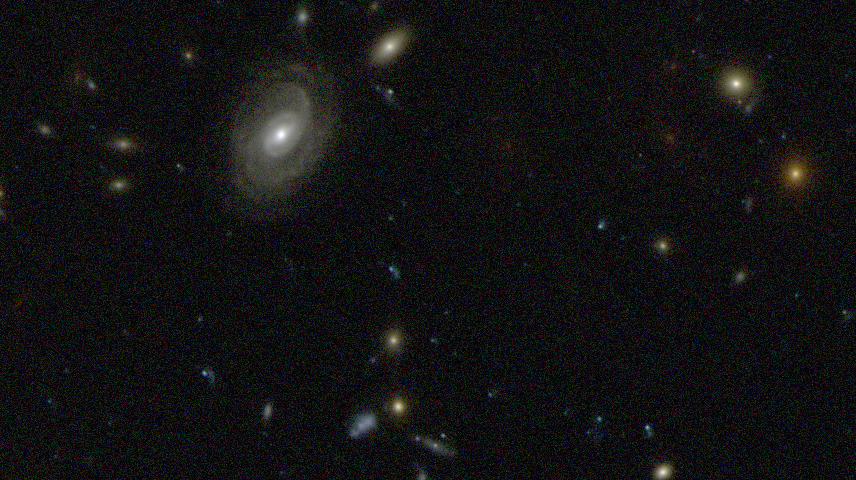}
  \caption {In the top two panels we show two galaxies used to produce
    the source plane. From left to right: the color composite image,
    the denoised image of the {\tt{F775W}} band, the original HXDF image
    (with the other objects present in the postage-stamp already removed) and
    the residuals. In the bottom panel we show a cut out of the source
    plane used to produce Fig.~\ref{fig_color}
    but without the addition of the substructures.}
  \label{fig:denoised}
\end{figure}

In Fig.~\ref{fig:denoised} we show and example with two denoised galaxies (color composite image and {\tt{F775W}} band), together with the original HXDF image and the residuals. The bottom panel of the same figure shows a zoom-in of the source plane, without the additional substructures, used to produce the image of Fig.~\ref{fig_color}.

\subsection{Lens model}
The lensing effects can be produced by any mass distribution, such as analytical dark matter halo models or numerical and/or hydrodynamical simulations. {In the examples shown here,} we use the simulated galaxy cluster \emph{Ares} \citep[which was extensively discussed in the paper by][]{meneghetti17}. It was  created by using the semi-analytic code {\tt MOKA}\footnote{\url{ http://cgiocoli.wordpress.com/research-interests/moka}}\citep{giocoli12}, and assuming a flat cosmology as described above.  {In short, {\tt MOKA} combines several mass components that make up the mass distribution of a galaxy cluster. These are a smooth dark matter halo, the sub-halos, and the brightest-central-galaxies (BCGs). Each of these components is fully parametrized and modeled consistently with the findings of state-of-the-art N-body and hydrodynamical numerical simulations.} From the three dimensional density distribution of the lens, {\tt MOKA} creates a projected two-dimensional surface density map. This map is used to calculate a field of deflection angles $\mathbf{\vec{\hat{\alpha}}}$, {which is one of the  inputs for \sky. Using the deflection angles, light rays} are traced back from the detector to the source plane through a grid of a given size in the lens plane.  {The mapping between the lens and the source planes is done using the lens equation. In the case of a single lens plane, this is written as
\begin{equation}
\label{eq:lens}
\vec{\beta}=\vec{\theta} - \frac{D_{\text{ls}}}{D_{\text{s}}}\vec{\hat{\alpha}}(\vec{\theta})
\end{equation}
In Equation \ref{eq:lens}, $\vec{\beta}$ is the position of the photon in the source plane, $\vec{\theta}$ is its apparent position on the detector, and $D_{\text{ls}}$ and $D_{\text{s}}$ represent the cosmology-dependent angular diameter distances between the lens and the source and between the observer and the source, respectively. Note that this formalism is general and valid for both the weak and strong lensing regimes. In the latter case, one position $\vec{\beta}$ can correspond to more than one coordinate $\vec{\theta}$.  

The deflection angle field can then be used to calculate other related and useful lensing quantities. For example, the convergence is half the divergence of the deflection angle:
\begin{equation}
\kappa(\vec\theta)=\frac{1}{2}\left[\frac{\partial \alpha_1}{\partial\theta_1}(\vec\theta)+\frac{\partial \alpha_2}{\partial\theta_2}(\vec\theta)\right] \;.
\label{eq:effconv}
\end{equation}

Shear, magnification (on the source and lens planes), as well as the locations of the caustic and critical lines are also readily derived from the deflection angles \citep[e.g.][]{meneghetti17}.}
\subsection{Multiple lens planes}
\label{sec:multiple}

\begin{figure*}
\centering
\resizebox{\hsize}{!}{\includegraphics{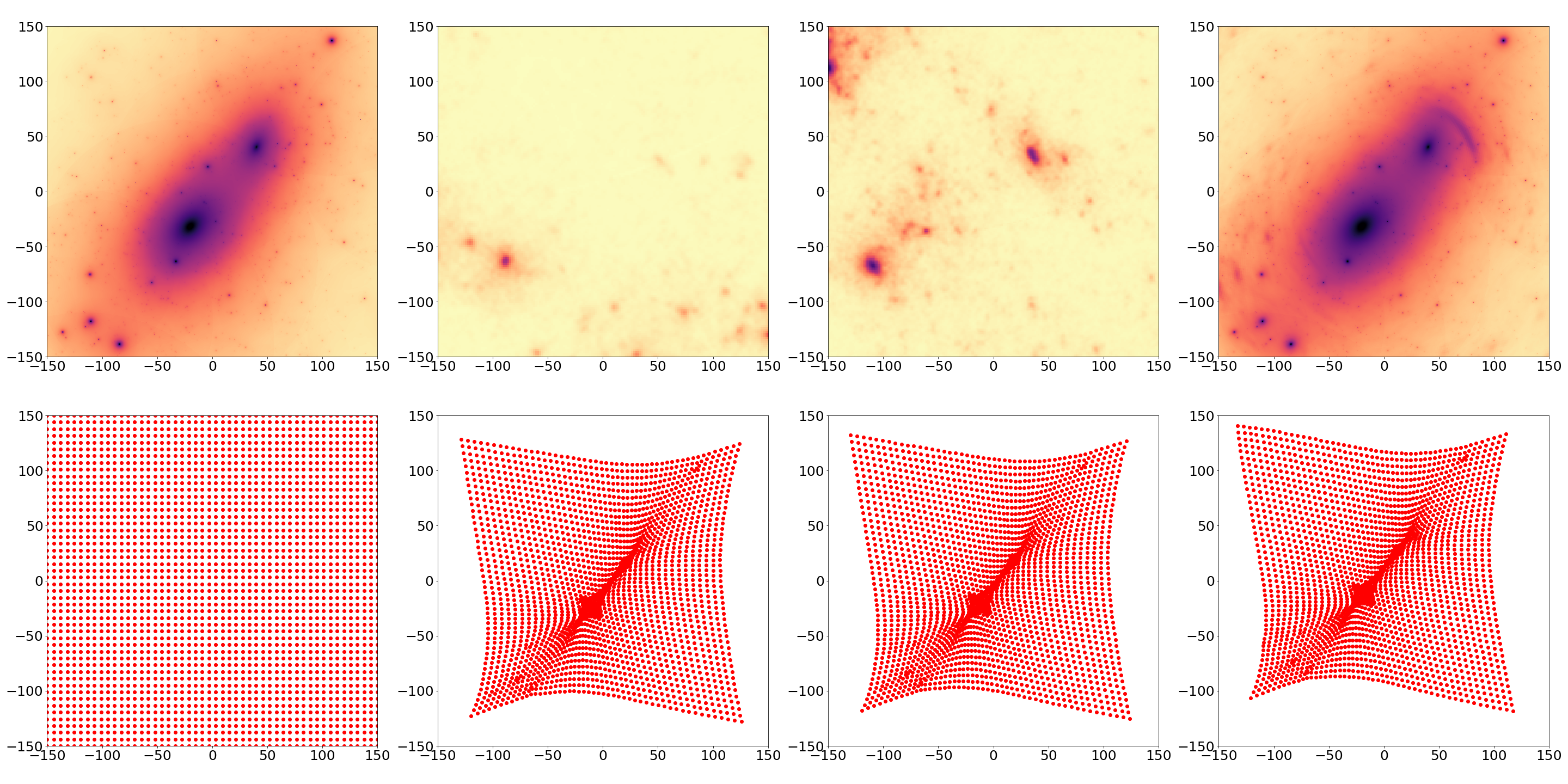}}
\caption{Example of multi-lens plane simulation. Upper panels: the left panel shows the convergence map of the {\em Ares} cluster. In second and in the third panels we show two mass sheets extracted from a cosmological simulation at $z=0.7$ and at $z=1$. We use the multi-plane ray-tracing technique to propagate light rays up to $z=12$ (fourth panel). The resulting effective deflection angles are used to compute a map of the effective convergence shown in the left panel. Bottom panels: in the leftmost panel, we show the regular grid through which we propagate the light rays on the first lens plane. In the other panels we show the results of applying the multi-plane lens equation up to $z_{\text{s}}=12$, including at each step the lens planes shown in the upper panels. The field-of-view in each panel is $150"\times 150"$. }
\label{fig:effconv}
\end{figure*}

{A major improvement of this version of \sky\ is the capability to simulate deflections caused by multiple lens planes \citep[e.g.][]{2014MNRAS.445.1954P}. When this feature is used, Eq.~\ref{eq:lens} is substituted by the lens equation in the form
\begin{equation}
\vec{\beta}=\vec\theta-\sum_{i=0}^{N_S}\frac{D_{\text{is}}}{D_{\text{s}}}\vec{\hat\alpha}^i(\vec\theta^i) \;.
\label{eq:multilens}
\end{equation}
In Eq.~\ref{eq:multilens}, $\theta=\theta^0$ is the position of the photon on the detector, here coinciding with the first lens plane; $\vec\theta^i$ and $\hat{\vec{\alpha}}^i(\vec \theta^i)$ are the photon position and the corresponding deflection angle on the $i-$th lens plane, respectively; and $D_{\text{is}}$ is the angular diameter distance between the $i-$th lens plane and the source. If the source falls behind $N_S$ lens planes all their contributions are accounted for in the sum on the right-hand side of Eq.~\ref{eq:multilens}.

By comparing the position of each photon on the first lens plane and in the source plane, we can define an {\em effective} deflection angle: 
\begin{equation}
\vec\alpha_{\text{eff}}=\vec\theta-\vec\beta=\sum_{i=0}^{N_S}\frac{D_{\text{is}}}{D_{\text{s}}}\vec{\hat\alpha}^i(\vec\theta^i) \;.
\label{eq:effdefl}
\end{equation}
In practice, this is the total deflection accumulated along the path between the observer and the source. It describes the effect of an effective mass distribution,  which can be derived using Eq.~\ref{eq:effconv} to compute the {effective} convergence, $\kappa_{\text{eff}}(\vec\theta)$.

In Fig.~\ref{fig:effconv}, we show an example of multi-lens plane simulation. In the upper-left panel, we show the surface density map of the cluster {\em Ares}. The cluster is at redshift $z=0.5$. We have extracted from a hydrodynamical simulation two slices of particles and we have used them to construct two additional lens planes at redshift $z=0.7$ and $z=1$. The details of the numerical simulation are in \citet{meneghetti10a}. The surface mass density on these two planes are shown in the second and in the third upper panels. Finally, we use the multi-lens plane formalism to compute the {\em effective} convergence for sources at $z=12$, which is shown in the upper-right panel. Note that, in the effective convergence map, the mass structures behind the cluster appear distorted by lensing. 

In the bottom left panel, we show an example of a grid of $50\times50$ points on the first lens plane, through which the light rays are propagated towards the sources. Before the rays are deflected, the grid is regular. The arrival positions of the light rays on the source plane at $z_{\text{s}}=12$ are shown in the other three bottom panels.  In each panel, we account for the deflections on an increasing number of lens planes: one, two, and all three lens planes, respectively. In all these cases, the grids are distorted and cover a smaller area than the original grid, due to the focusing effect of the lens. Most of the deflection comes from the cluster (which is here placed on the first lens plane). Structures along the line of sight cause minor shifts of the arrival position of the light rays on the source plane. Particularly evident is the effect of a mass clump located near the upper-right edge of the third lens plane, which causes deflections of several arc-seconds in that area.

\subsection{Multiple source planes}
\sky\ also implements multiple source planes. The sources extracted from the XUDF are divided in 100 redshift bins between $z=0$ and $z=12$. The bin sizes are defined such that their centers are equally spaced in lensing distance, $D_{\text{ls}}D_{\text{l}}/D_{\text{s}}$. We define one source plane for each redshift bin. The sources in the bin are placed on the corresponding plane and  are distorted using Eq.~\ref{eq:lens} or Eq.~\ref{eq:multilens}, depending on whether a single or multiple lens planes are used. This method results in a significant reduction of the computational time, compared to computing the deflections for each individual source redshift.
} 

\subsection{Substructures in source images}
\subsubsection{Morphological analysis of nearby galaxies}
Galaxy clusters act as natural telescopes that magnify faint source galaxies, effectively increasing the native resolution of the instrument at hand if the position of the source is sufficiently close to a caustic curve. This allows to resolve substructures that would not be otherwise visible in the absence of the lens. We include these features---which represent, for example, regions of active star formation---in the denoised HXDF postage stamps that \sky\ uses as input by modeling them as S\'ersic profiles of index $n=3.5$.\footnote{\textcolor{black}{The S\'ersic index of the knots of structure formation is a parameter that can be adjusted in the code.}} We use information inferred from a morphological analysis of nearby galaxies to derive empirically-motivated size and luminosity distributions for these substructures. \textcolor{black}{This assumes that the information regarding the shape of the size and luminosity distributions extracted from this local sample will be applicable to star-forming regions of high-redshift galaxies, which might not be true in general. However, the size and luminosity distributions of these regions are expected to follow power-law functions \citep{kennicutt89,johnson17a,johnson17b}, consistent with the results presented in this section and described below.}

We use the sample of nearby, well-resolved, multiband (\emph{g}, \emph{r}, and \emph{i} filters from the Thuan-Gunn photometric system \citep{thuan76,wade79,schneider83}), and ground-based galaxy images compiled by \citet{frei96}. The complete catalog by \citet{frei96} has a total of 113 galaxies that span all different Hubble classification classes. We use the morphological parameter {{\tt{T}}} provided by the catalog\footnote{In turn, \citet{frei96} report {{\tt{T}}} from the galaxy catalog by \citet{devaucouleurs91}.} to select a subsample of spiral galaxies, for which $0 \leq {{\tt{T}}} \leq 9$ is satisfied. Furthermore, by visual inspection of the remaining galaxies, we select those galaxies that are viewed completely or nearly face-on. The final sub-sample of images analyzed consists of 32 galaxies. 
\begin{figure}
\centering
\resizebox{\hsize}{!}{\includegraphics{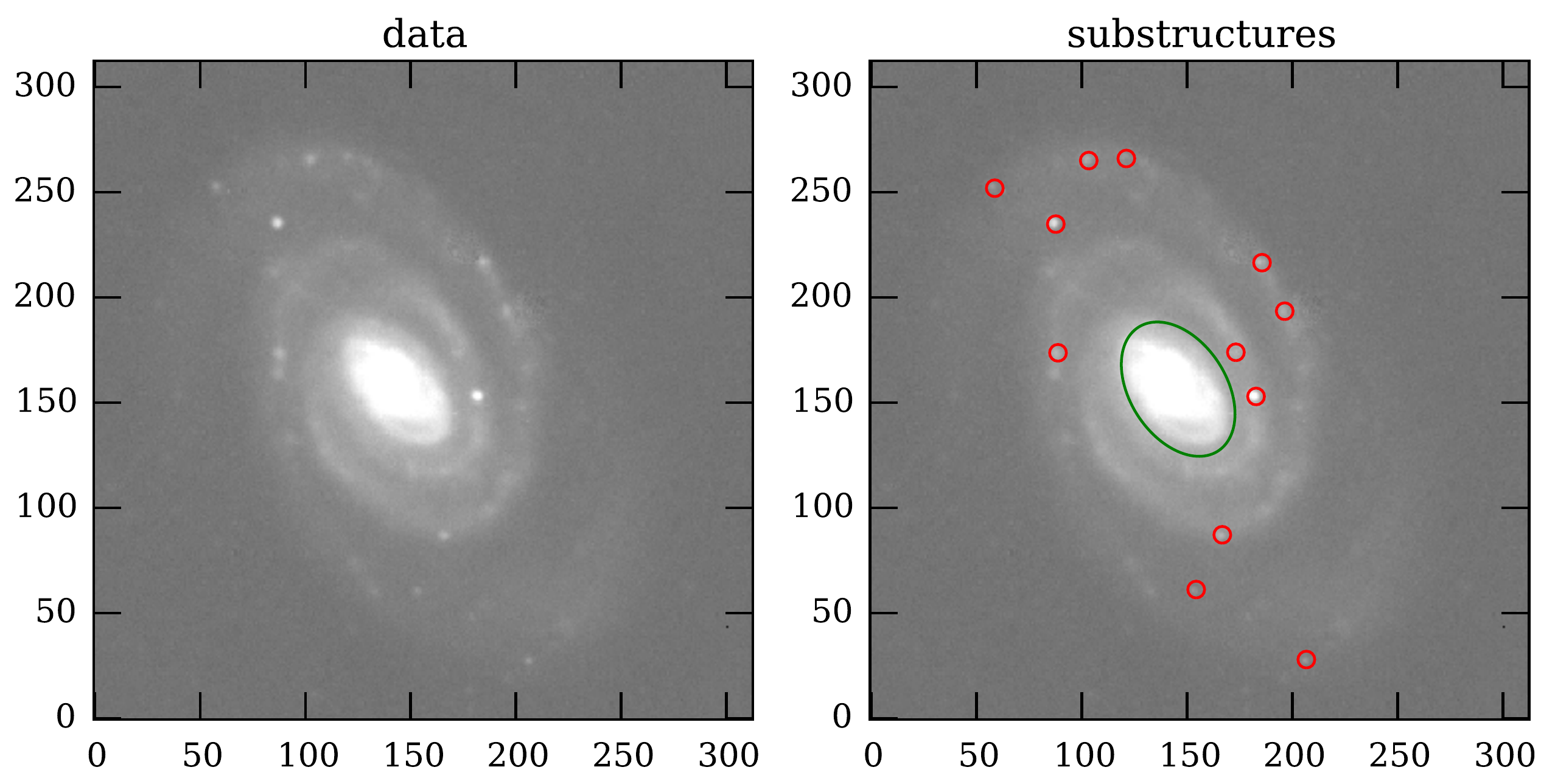}}
\caption{\emph{Left}: Postage stamp image of galaxy NGC 5364 in the \emph{r} band obtained from the catalog in \citet{frei96}. \emph{Right}: Following the procedure explained in the text, we use {\tt{SExtractor}} to detect substructures in the disk component of the galaxy (red circles), excluding most of the bulge region in the center (green ellipse).}
\label{f1}
\end{figure}
Each galaxy image is available as a postage stamp FITS file in each photometric band.\footnote{\url{zsolt-frei.net/catalog.htm}} The images are reduced and their instrumental signatures removed, and foreground stars have been identified and removed by means of an empirical PSF model \citep{frei96b}. 

To find the substructure regions in each image, we begin by smoothing the postage stamp with a Gaussian kernel with a typical size of $3.5$\textendash$10$ pixels, depending on the galaxy. The smoothed image is then subtracted from the original stamp, and the difference image is used as an input to {\tt{SExtractor}} \citep{bertin06}. We also define an elliptical bad-pixel mask that excludes the central region of the galaxy. To do this, we use the best-fit parameters obtained by using the code {{\tt{Lensed}}} \citep{tessore16} to fit the galaxy data to a bulge-plus-disk model, where each component is assumed to follow a S\'ersic profile. This bad pixel mask is also passed to {\tt{SExtractor}} as input. For each galaxy image, we modify relevant parameters\footnote{Such as {\tt{DETECT\_MINAREA}}, {\tt{DETECT\_MAXAREA}}, and {\tt{DETECT\_THRESH}}.} in the {\tt{SExtractor}} configuration file to \textcolor{black}{customize} the detection of the substructure clumps \textcolor{black}{in each image}. {\tt{SExtractor}} then outputs a catalog that includes pixel positions in pixels, half-light radii in pixels, and flux (in ADU) parameters ({\tt{XWIN}}, {\tt{YWIN}}, {\tt{FLUX\_RADIUS}}, and {\tt{FLUX\_AUTO}}, respectively). At the catalog level, we also perform a selection that excludes objects that are either too small or too bright, rejecting outliers in the size and flux distributions by means of $3\sigma$-clipping. Fig.~\ref{f1} shows an example of the type of substructures that are detected, along with the original image from the \citet{frei96} catalog. 
\begin{figure}
\centering
\resizebox{\hsize}{!}{\includegraphics{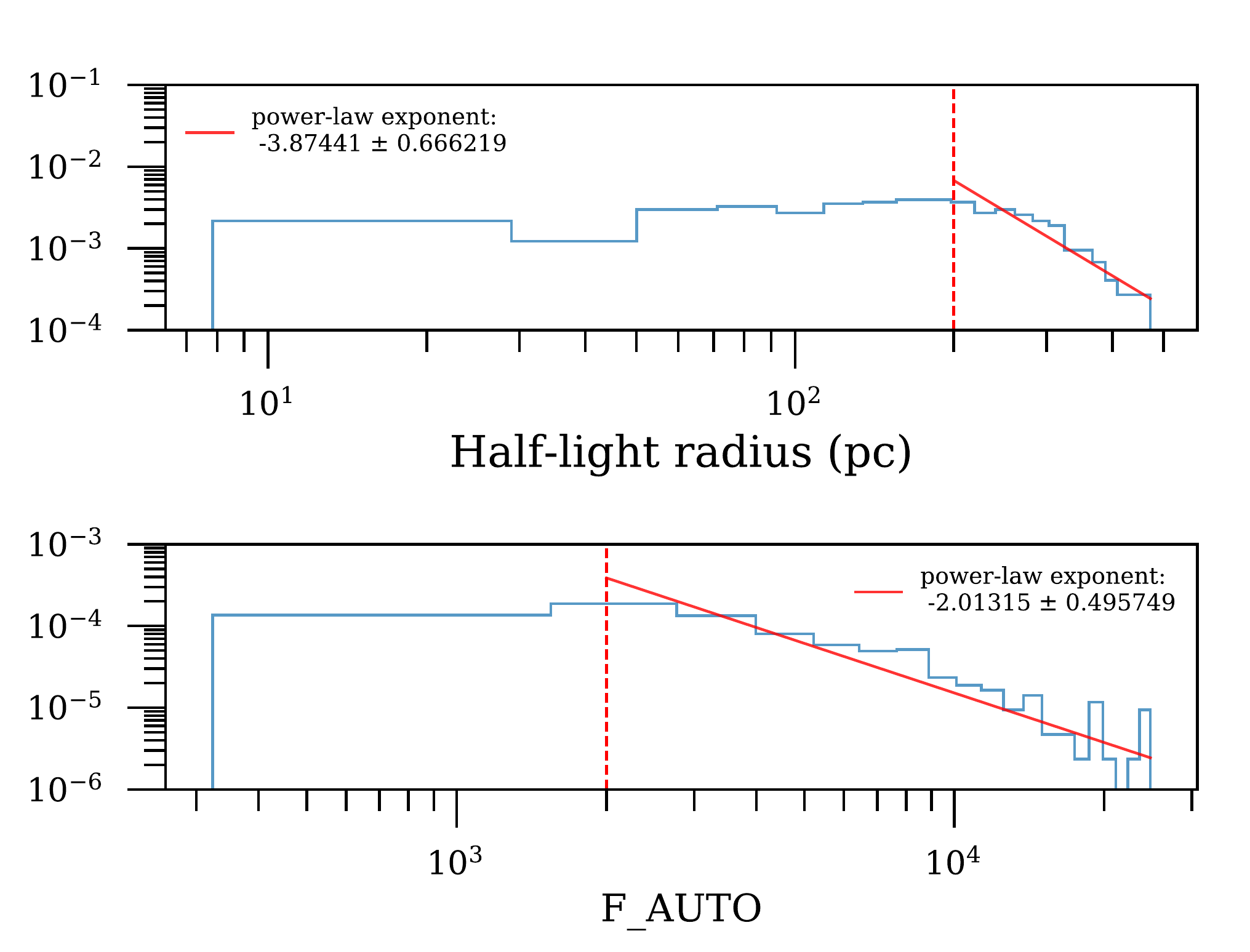}}
\caption{Size and luminosity distributions of substructures identified from the galaxies in \citet{frei96} (arbitrary amplitudes). We fit a power law to the tail of the distribution (to mitigate selection biases at the lower end of both distributions), and use the best-fit exponents to assign a size and a luminosity to every S\'ersic substructure created by \sky\ in each source galaxy postage stamp, as described in the text.}
\label{f2}
\end{figure}

Once we have a catalog of candidate structures (from a total of 350), we calculate their physical size in parsecs from their half-light radii in pixels by using the angular diameter distance to the galaxy that hosts each substructure. We assume a flat cosmology of the form specified above, and use the mean redshift to each galaxy in our subsample published by different sources as compiled by the NASA/IPAC Extragalactic Database.\footnote{\url{https://ned.ipac.caltech.edu}} 

Fig.~\ref{f2} shows the final flux and size distribution obtained. We fit the tail of each distribution to a power law (\textcolor{black}{to avoid selection biases at the lower end of the distribution}), and obtain the best-fit exponents $-2.01 \pm 0.49$ and $-3.8 \pm 0.68$, respectively, and use these results as motivation to choose a power-law flux and size distributions of exponents $-2$ and $-4$ when creating substructures in each source galaxy used by \sky. However, these parameters can be adjusted in the code, as well as the choice of the functional form for their probability distributions. 

\subsubsection{Adding substructures to the source galaxy images}
\label{sect:subadd}

For a given source galaxy image, we select a fraction $Q$ of its total flux to be redistributed in the form of substructures. We assign a different fraction $Q$ depending on the morphological classification of each galaxy, determined by their assigned \citep{coe12} interpolated SED template (11 basis templates), ranging from $N=1$ for early-type galaxies to $N=11$ for starburst galaxies. Thus, we set $Q$ to 0.05 for lenticular galaxies, $Q$ to a random number drawn from a uniform distribution in the interval [0.2,0.4], and 0 otherwise (\textcolor{black}{i.e., late-type galaxies should exhibit more regions of star formation while no substructures are created for early-type galaxies}). As with other parameters, $Q$ can be adjusted in \sky\ if desired. %{\bf MM: do we have some reference supporting this choices of the parameter Q?}
\begin{figure}
\centering
\begin{tabular}{|c|c|}
%\hline
\subf{\includegraphics[width=30mm]{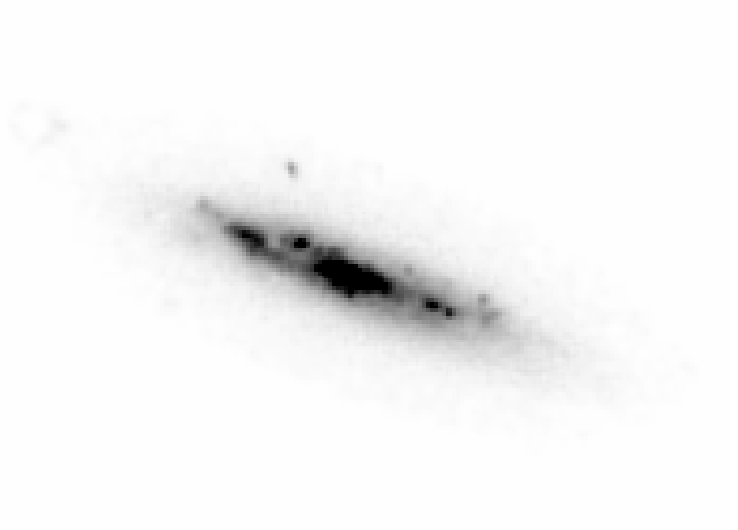}}
     {}
&
\subf{\includegraphics[width=30mm]{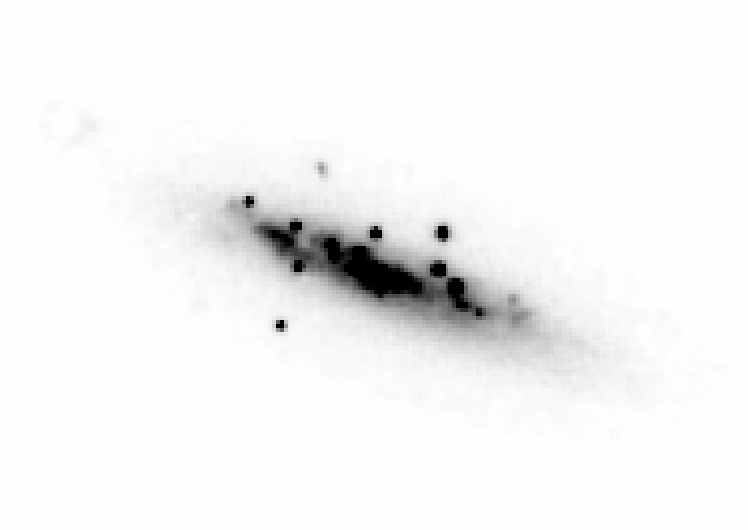}}
     {}
\\
%\hline
\subf{\includegraphics[width=30mm]{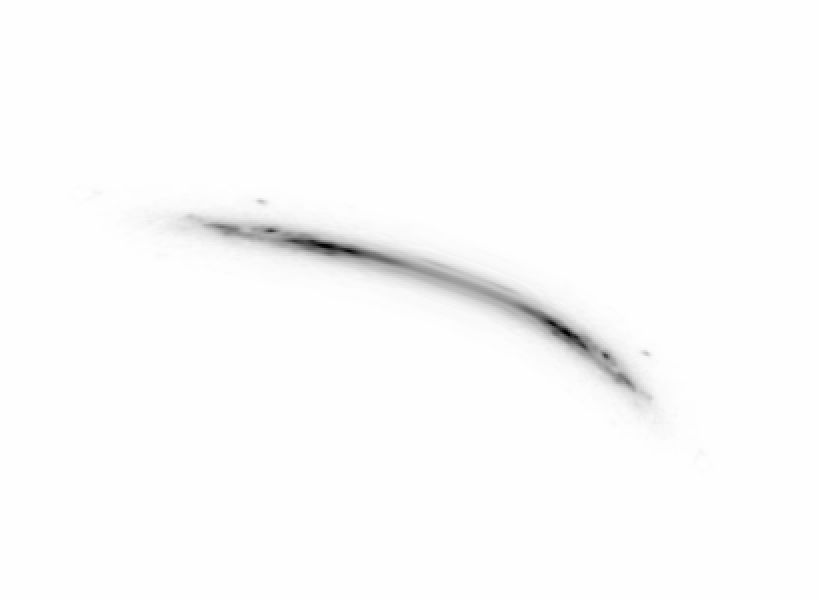}}
     {}
&
\subf{\includegraphics[width=30mm]{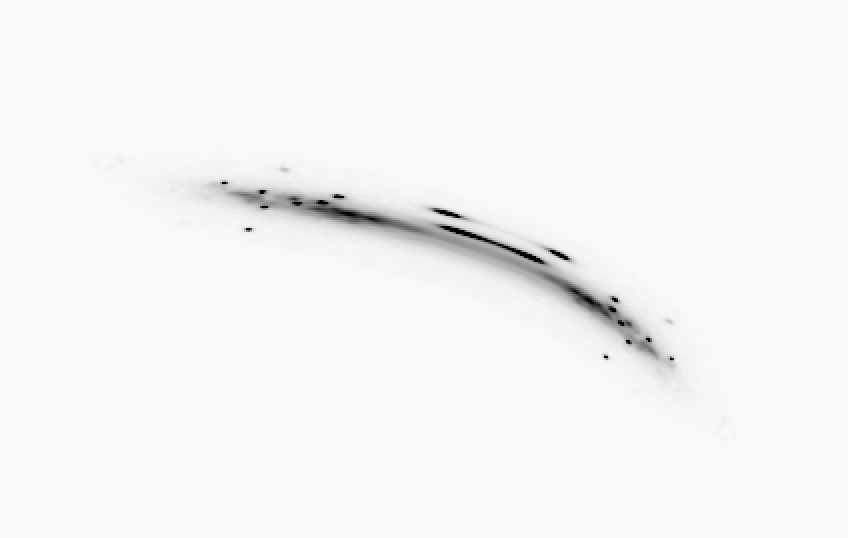}}
     {}
\\
%\hline
\end{tabular}
\caption{Galaxy from the \citet{rafelski15} catalog (ID: 22245). The upper row shows the unlensed galaxy with and without knots of substructure (left and right panels). The lower row shows a similar comparison after the galaxy has been strongly lensed.}
\label{fig_with_without_sub}
\end{figure}
Each substructure is placed at a pixel location derived by creating a histogram of the surface brightness of postage stamp by means of an inverse transform sampling algorithm. In this way, the substructures will be placed mainly in locations that are proportional to the surface brightness distribution of the source galaxy. Fig.~\ref{fig_with_without_sub} shows an example of a source galaxy with and without substructures. Once the location of a particular substructure has been chosen, we model it as a S\'ersic profile of index $n=3.5$, with a particular size and luminosity drawn from the distributions of Fig.~\ref{f2}. These substructures will be subsequently created only if the image of their host galaxy in the lens plane is close enough to a critical curve such that the local magnification produced by the lens is larger than a certain threshold $\mu_{\mathrm{t}}$. 

%We note that this method finds a discrete pixel position for the center of the substructure knots, assigning one substructure per pixel. In principle, however, more than one knot could land in a single pixel. One way to account for this in a future version of the code would be to randomize the position of the knot within the chosen pixel, allowing for the possibility that, in that case, the flux of the substructure might be distributed among more that one pixel if its centroid falls near one of the edges. 

\subsection{Virtual observations}
As explained above, \sky\ has the capability of producing a virtual observation with any given instrument and/or telescope and at any desired resolution, for a particular field of view. 

{Once the telescope (and the detector) and the lens (or lenses) have been defined, the code reconstructs the images of the sources by assigning to the pixels on the detector, whose positions are given by the $\vec \theta$ vectors, the value of surface brightness at positions $\vec\beta$,
\begin{equation}
I(\vec\theta)=I_{\text{s}}(\vec\beta) \;, 
\end{equation}
with $\vec\beta$ given either by Eq.~\ref{eq:lens} or by Eq.~\ref{eq:multilens}, thus accounting for the lensing effects by the matter along the line of sight.
}

Finally, the images are convolved by the instrumental Point Spread Function (PSF), and different sources of noise such as sky background, Poisson, and readout noise are added, depending on the specified number of exposures as explained in Sect.~\ref{sect:genmeth}.  
 
\textcolor{black}{\subsection{Code validation}
The different components of the \sky\ pipeline have been validated independently.  \citet{maturi16} use simulations to corroborate the quality of the denoised source images in terms of moments of surface brightness. \citet{vanzella17} show in their appendix that the simulated observations created by \sky\ are able to produce realistic background levels. \\
The ray-tracing algorithm has been validated in previous version of the \sky. In addition, we use the image of local dwarf galaxy NGC1705 (obtained by the LEGUS survey\footnote{\url{https://archive.stsci.edu/prepds/legus/dataproducts-public.html}} in the F275W band) to further validate this part of the code. The left panel of Fig.~\ref{fig:skylens_test} shows an unlensed image of the galaxy, with the positions of a number of star clusters marked by circles. The galaxy has been placed at z=6.14 and lensed by \sky\ using the lens model produced by \citet{caminha17} for the Frontier Field MACS0416. The lensed image (PSF free) is shown in the right panel of the figure, where the lensed positions of the star clusters shown in the left panel---as expected by the public software {\tt{Lenstool}} \footnote{\url{https://projets.lam.fr/projects/lenstool/wiki}}---are marked again with circles. It can be seen that the circles enclose the lensed star clusters in the \sky\ simulated image. In addition, the positions of these knots also reflect the inversed parity of the image. In this particular case, we have confirmed that the position of the galaxy image in the source plane with respect to the caustics of the system is such that it is lensed into three images. The lensed image shown in the right panel of Fig.~\ref{fig:skylens_test} lies inside the resulting critical lines of the lens plane, with a negative parity. \\
\begin{figure}
\centering
\resizebox{\hsize}{!}{\includegraphics{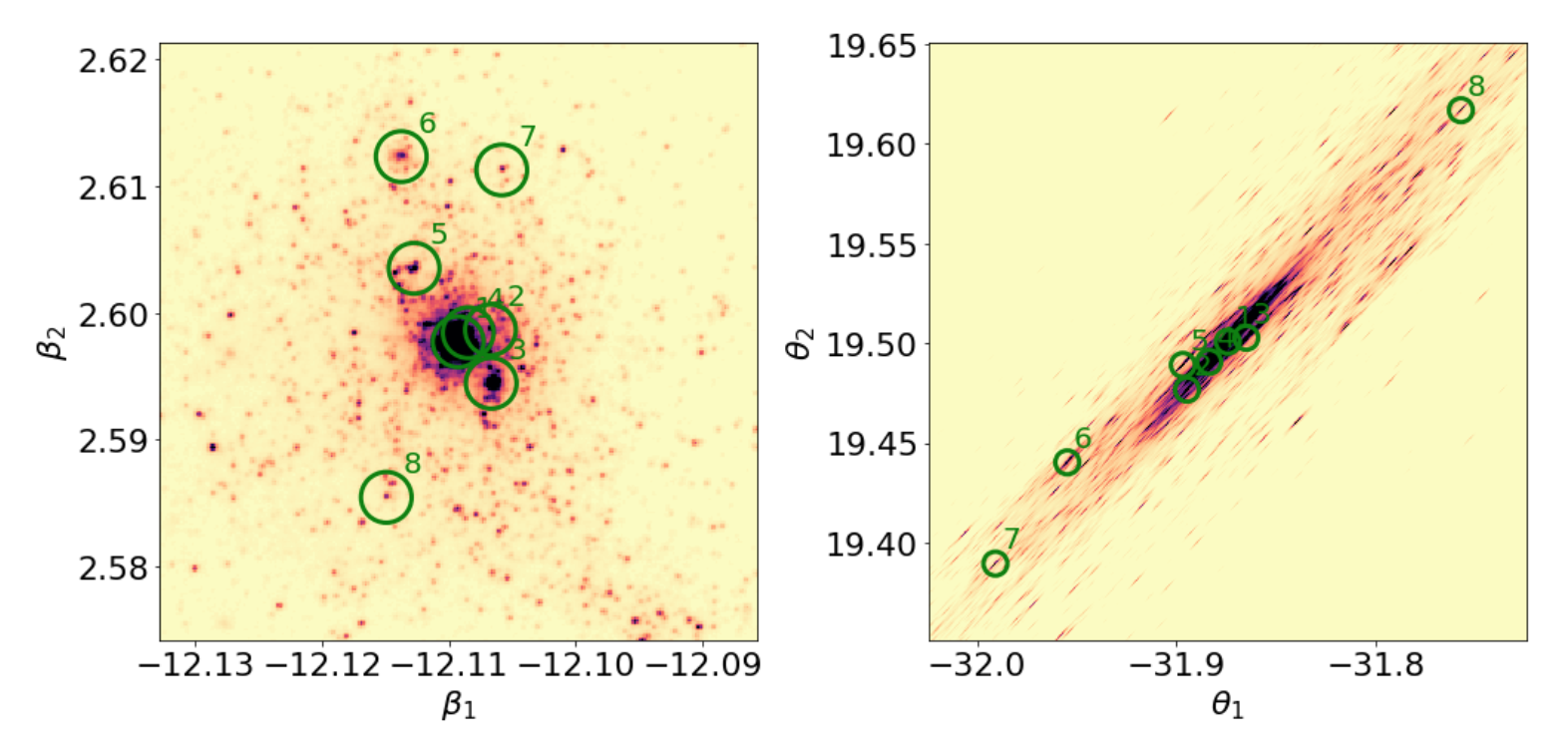}}
\caption{\emph{Left}: Unlensed image of the galaxy NGC1705 (LEGUS survey, F275W band). The circles mark the position of star clusters within the galaxy.\emph{Right}: the source is placed at z=6.14 and lensed with \sky\, using the lens model of the Frontiers Field MACS0416 by \citet{caminha17}. The circles mark the positions of the lensed star clusters, as expected by the software {\tt{Lenstool}}. The parity of the lensed image is negative, consistent with the position of the source with respect to the caustics and with the lensed image itself with respect to the critical lines in the lens plane.}
\label{fig:skylens_test}
\end{figure}
The formalism of gravitational lensing in multiple planes described in section~\ref{sec:multiple} is standard and has been implemented and validated by other lensing simulation codes such as \citet{giocoli12,birrer18,metcalf18}. In particular, we have verified that \citet{birrer18} produces results consistent with those of \sky\ as shown in Fig. \ref{fig:effconv}.} 

%1) sources: in my paper I described the quality of the denoised images in terms  
%of moment of brightness. I would just mention that
%2) lensing:
%- ray-tracer: Massimo, I guess we can just refer to some of your old  
%papers, otherwise...
%-Background, image quality.  Able to reproduce the right background level in images: Vanzella simulated observations of sources 
%-Ray tracing: previous versions can reproduce location of multiple images
%-Multiple planes formalism: MOKA, Carlo created light cones, formalism is identical. Implementation in GLAMER
%-comparison between multiple planes based on standards tested by simon berrier and B. Metcalf 

\section{Lensing simulations}
\label{sec:sims}
We have simulated observations of galaxy clusters through five different telescopes and instruments to illustrate the capabilities of \sky: \emph{HST} ACS WFC, \emph{WFIRST} WFI, \emph{JWST} NIRCam, Subaru HSC, and {\em Euclid} VIS. In the first four cases, we use the corresponding PSF-generating tools to produce sensible PSF models for different filters, and the Exposure Time Calculators to obtain estimates of the sky background. For the latter, we refer to the {\em Euclid} red-book \citep{2011arXiv1110.3193L} to generate a simple PSF model. We use the HXDF source catalog and the {\tt{ARES}} lens model  described in Sections 2.1 and 2.2 above. 
\begin{figure*}
\centering
\begin{tabular}{|c|c|}
%\hline
\subf{\includegraphics[width=80mm]{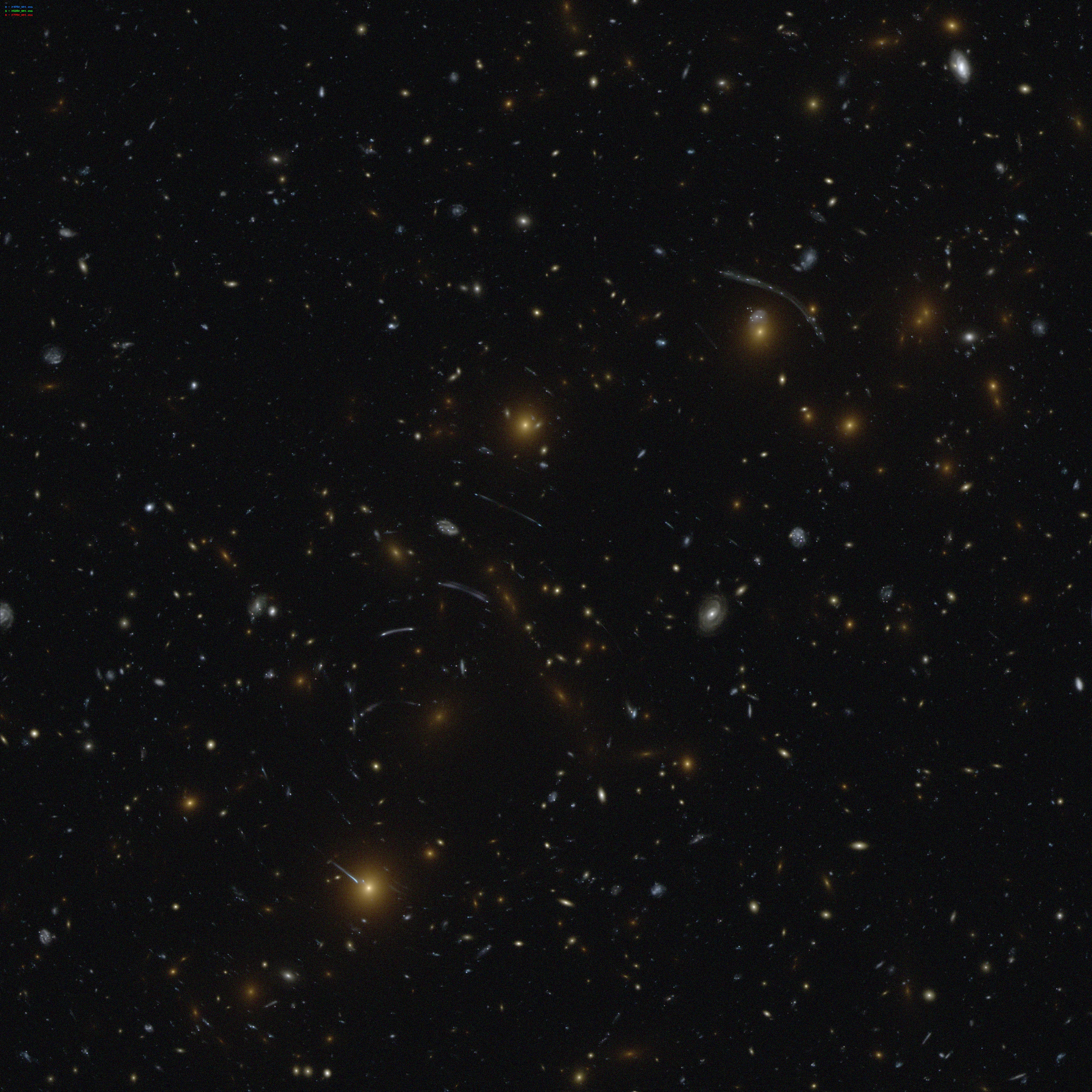}}
     {}
&
\subf{\includegraphics[width=80mm]{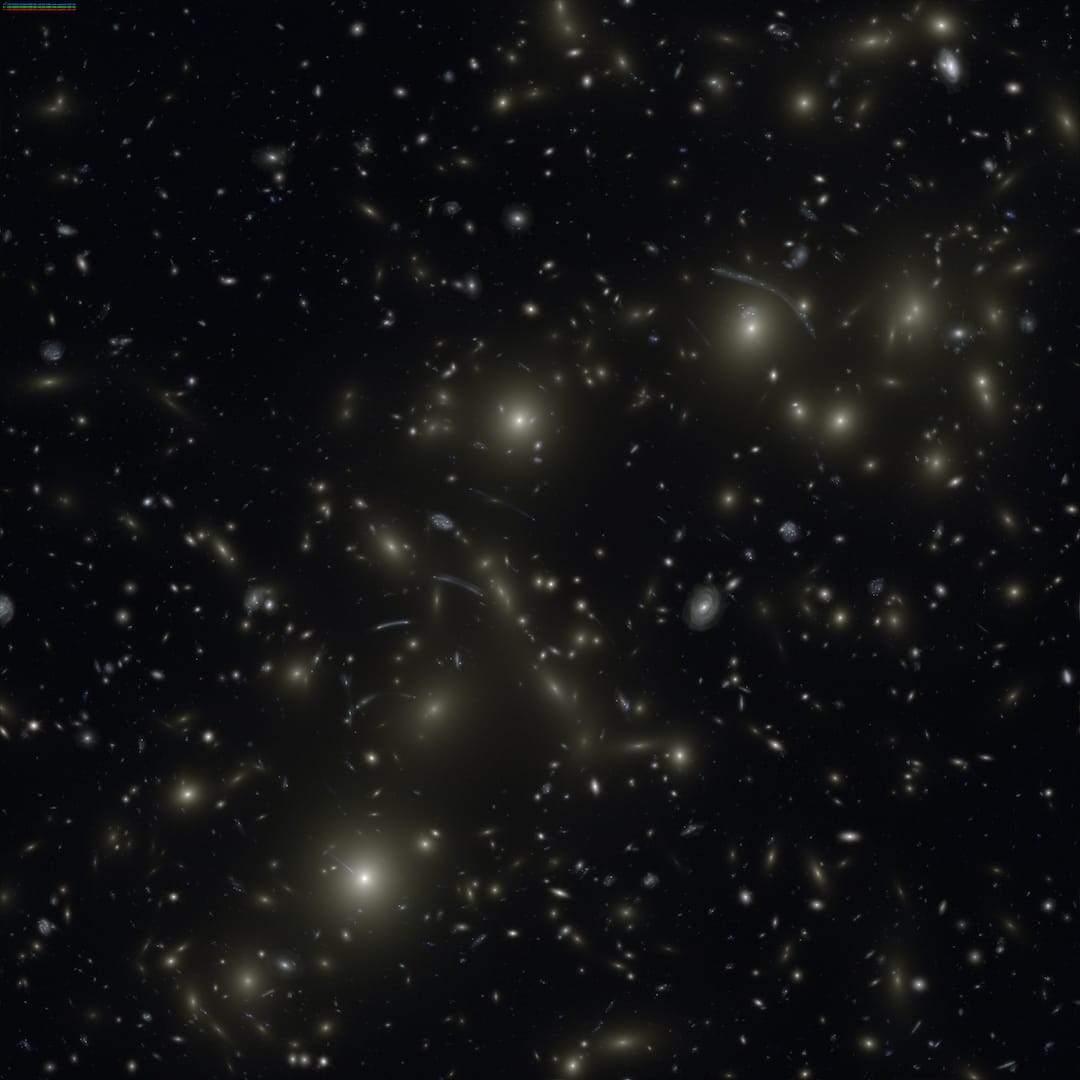}}
     {}
\\
%\hline
\subf{\includegraphics[width=80mm]{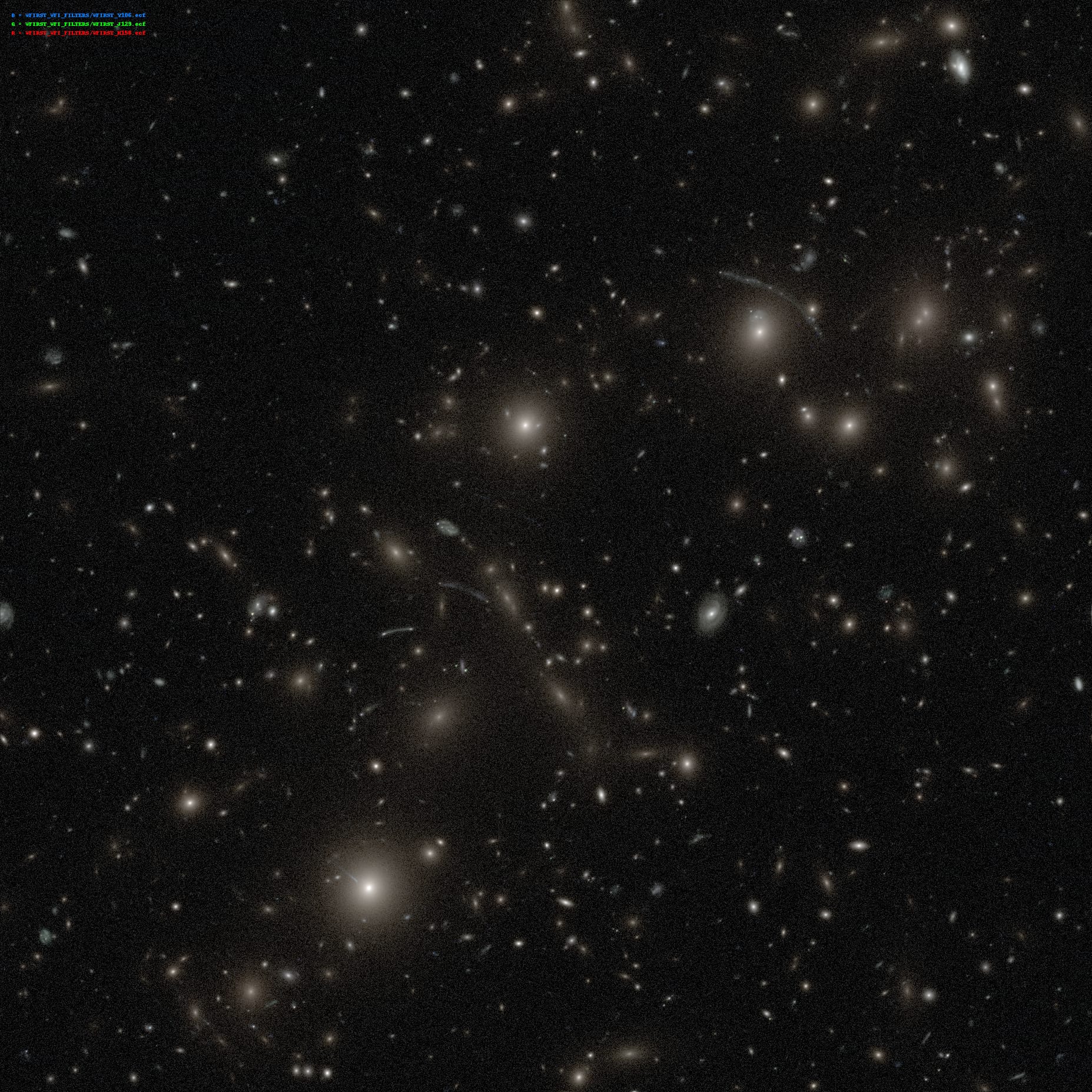}}
     {}
&
\subf{\includegraphics[width=80mm]{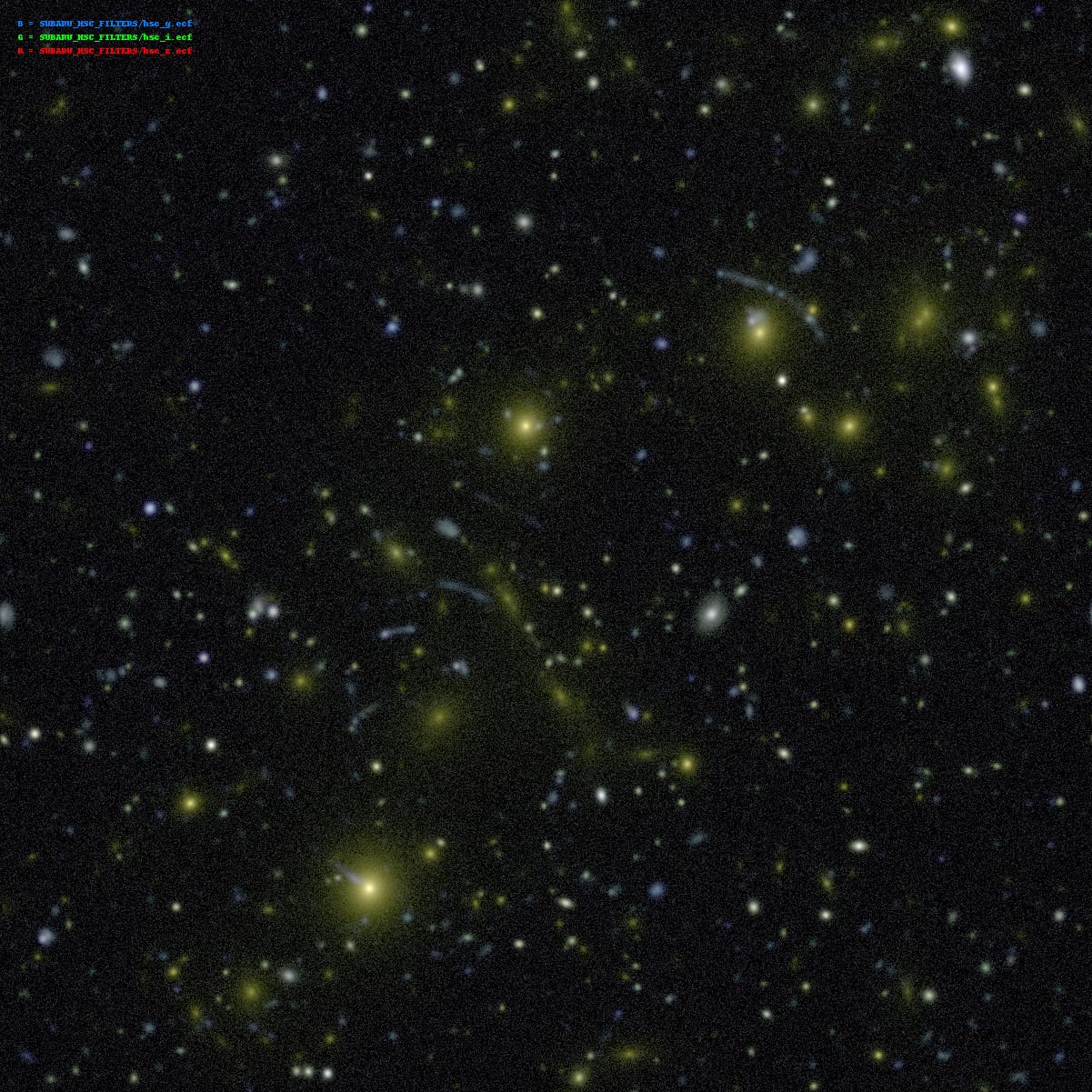}}
     {}
\\
%\hline
\end{tabular}
\caption{Color composites from images produced by \sky. In all case the FOV is $204''\times 204''$.\emph{Upper left}: \emph{HST} ACS WFC, produced by combining images in the {\tt{F475W}}, {\tt{F625W}}, and {\tt{F775W}} filters. The exposure time for each image is 5000 s. \emph{Upper right}: \emph{JWST} NIRCam, produced by combining images in the {\tt{F090W}}, {\tt{F150W}}, and {\tt{F200W}} filters. The exposure time for each image is 10146 s. \emph{Lower left}: \emph{WFIRST} WFI, produced by combining images in the  {\tt{Y106}}, {\tt{J129}}, and {\tt{H158}} filters. The exposure time for each image is 504 s. \emph{Lower right}: Subaru HSC, produced by combining images in the  {\tt{g}}, {\tt{i}}, and {\tt{z}} filters. The exposure time for each image is 600 s.}
\label{fig_color}
\end{figure*}
%originals: HST_ACS_COLOR_WITHCL.png
%JWST_NIRCAM_COLOR_WITHCL_2.png
%WFIRST_WFI_COLOR_WITHCL.png
%SUBARU_HSC_COLOR_WITHCL.png
We show the resulting images in Fig. \ref{fig_color} (each one with a  field of view of $204'' \times 204''$), where we have combined three images in the ``red", ``green", and ``blue" channels by using the software {\tt{trilogy}} \footnote{\url{http://www.stsci.edu/~dcoe/trilogy/Intro.html}} \citep{coe12}. Fig. \ref{f5} shows a zoomed-in region of the panels in Fig. \ref{fig_color} illustrating a strongly-lensed arc with the star-forming regions added as described in Section~\ref{sect:subadd}. 

\subsection{\emph{HST} ACS WFC}
We simulate observations through the ACS WFC of \emph{HST} in the {\tt{F475W}}, {\tt{F625W}}, and {\tt{F775W}} filters, each one with an exposure time of 5000s. We use the web interface\footnote{\url{http://tinytim.stsci.edu/cgi-bin/tinytimweb.cgi}} of the {\tt{Tiny Tim}} software \citep{krist11} to generate the PSF models in each band. The PSF models have a pixel scale of $0.0495''$ per pixel, but the final images will be rendered at a resolution of $0.03''$ per pixel. We obtain values for the sky background (in e$^-$/pix/sec) from the measurements performed by \citet{sokol12} (Table 2, average backgrounds). 

%F435W: 36.4/ 1000  = 0.0364
%F475W: 62.1/ 1000 = 0.0621
%F606W: 132.5/ 1000 = 0.1325
%F625W: 87.7 / 1000 = 0.0877
%F775W: 83.1 / 1000 = 0.0831
%F814W: 108.0 /1000 = 0.108
%F850LP: 45.3/1000 = 0.0453

\begin{figure}
\centering
\resizebox{\hsize}{!}{\includegraphics{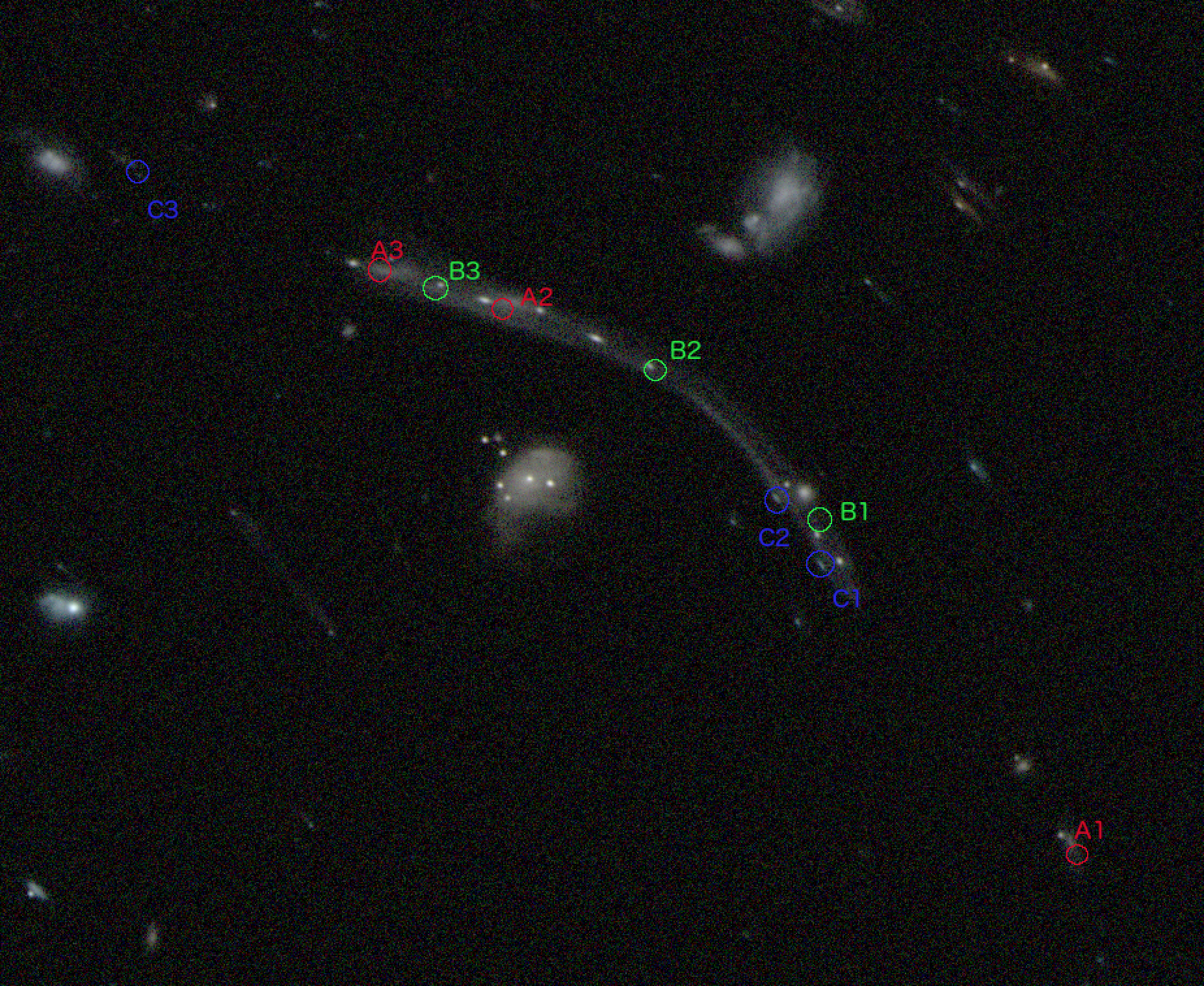}}
\caption{Zoomed-in of the upper central region in the \emph{HST} ACS WFC image of Fig.~\ref{fig_color}. The approximate position of the multiple images from three background sources (labeled as A,B, and C) have been identified. The lensed arc is a combination of multiple lensed sources. Multiple images of knots of substructure can also be seen in the arc.}
\label{f5}
\end{figure}

\subsection{\emph{JWST} NIRcam}

We simulate observations in the imaging mode of the \emph{JWST} NIRCam through the short-wavelength channel (0.6-2.3 $\mu$m) in the {\tt{F090W}}, {\tt{F150W}}, and {\tt{F200W}} filters. We use the \emph{JWST} PSF simulation tool {\tt{WebbPSF}}\footnote{\url{https://jwst.stsci.edu/science-planning/proposal-planning-toolbox/psf-simulation-tool-webbpsf}} to produce PSF models in each filter. The native pixel scale of the instrument is $0.032''$ per pixel, however, the NIRCam short wavelength is undersampled at 2.4 microns. Thus, we created PSF models sampled at a scale of 0.032/N with N=4 to satisfy the Nyquist criterium (2p/$\lambda_m$F, which results in a number of dithered exposures N $\geq$ 3 for and f-number F of 20, a pixel pitch p of 18 $\mu$m and minimum wavelength $\lambda_m$ of 0.6 $\mu$m).  We estimate the contribution from the background in each filter from four exposures by using the \emph{JWST} Exposure Time Calculator\footnote{\url{https://jwst.etc.stsci.edu/}} for NIRCam in the {\tt{DEEP8}} readout pattern with 10 groups. The values recorded for each of the 3 filters used are 0.318, 0.321, and 0.289 e$^-$/pix/s, respectively. The choice of the readout pattern ({\tt{DEEP8}}) and the number of exposures (N=4) sets the exposure time of each image to 10146 s. 

\subsection{WFIRST} 

We produce simulations in three of the four bands of the planned High Latitude Survey by the Wide Field Imager of \emph{WFIRST}: {\tt{Y106}}, {\tt{J129}}, and {\tt{H158}}. We use the \emph{WFIRST} module developed by \citet{kannawadi15} to obtain PSF models and sky background levels in each band.\footnote{We use {\tt{GalSim v1.4}}. The \emph{WFIRST} module is called {\tt{galsim.wfirst}}, and the PSF and sky backgrounds are obtained by calling the utilities {\tt{galsim.wfirst.getBandPasses()}} and {\tt{galsim.wfirst.getSkylevel()}}, respectively.} The near-infrared detectors of the WFI have a native pixel scale of $0.11''$ per pixel; however we draw them at a scale of $0.11''$/N per pixels with N=3 to avoid undersampling. The sky background model reported by {\tt{GalSim}} includes zodiacal light, stray light (10\%), and thermal backgrounds, and makes use of the Exposure Time Calculator\footnote{\url{http://www.tapir.caltech.edu/\~chirata/web/software/space-etc/}} by \citet{hirata12}. Including a mean dark current of 0.0015 e$^-$/pix/s, the sky backgrounds obtained for the {\tt{Y106}}, {\tt{J129}}, {\tt{H158}}, bands are 0.669, 0.654, and 0.654, e$^-$/pix/s respectively. The exposure time chosen for each image is 504 s (168 s per exposure).

\subsection{HSC}

We use the ``PSF Picker" tool to generate PSF models in the {\tt{g}}, {\tt{i}}, and {\tt{z}} bands of the HSC survey for the Wide Field Survey described in \citet{aihara17}.\footnote{The tool can be found at: \url{https://hsc-release.mtk.nao.ac.jp/psf/pdr1/}} The parameters used for the query are (RA, DEC)=(180.0, 0.0) deg., tract 9348, patch ``8,8", coadd, and pdr1\_wide., which represent a location in the wide-survey area. The pixel scale of the PSF models is $0.17''$ per pixel, and the seeing values were set to $0.72''$, $0.56''$, and $0.63''$ in each of the {\tt{g}}, {\tt{i}}, and {\tt{z}} bands, respectively, using the values in \citet{aihara17}. We use the HSC Exposure Time Calculator\footnote{\url{https://hscq.naoj.hawaii.edu/cgi-bin/HSC_ETC/hsc_etc.cgi}} to estimate the contributions due to the sky backgrounds. Under the conditions of gray time (7 days after new Moon), transparency of 0.9, and 60 degrees of separation from the Moon, we obtain sky values of 35.08, 75.74, and 45.60 e$^-$/pix/s for the {\tt{g}}, {\tt{i}}, and {\tt{z}} bands, respectively. The exposure time for each band was set to 600 s.  

\begin{figure*}
\includegraphics[width=0.49\hsize]{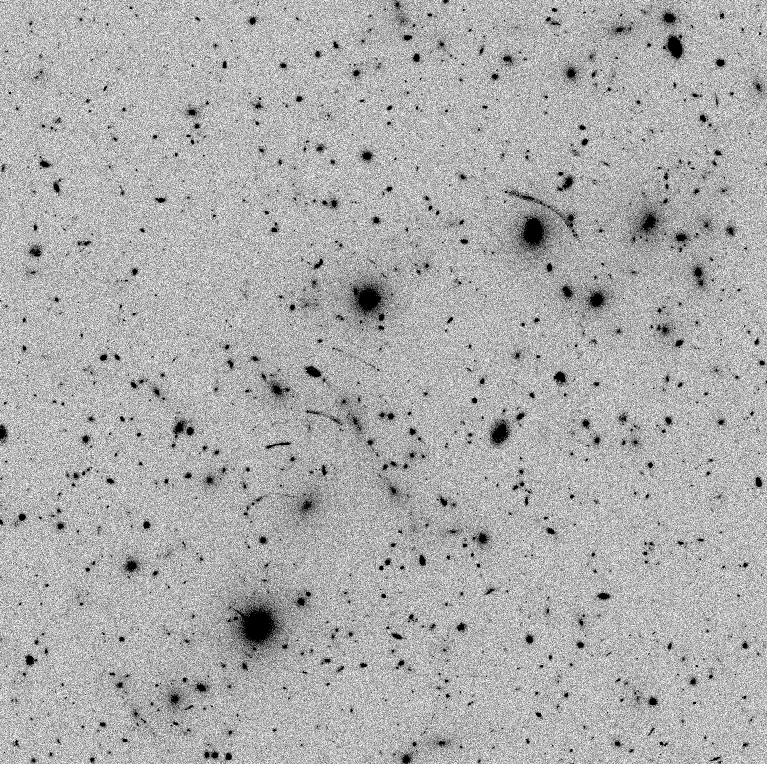}
\includegraphics[width=0.49\hsize]{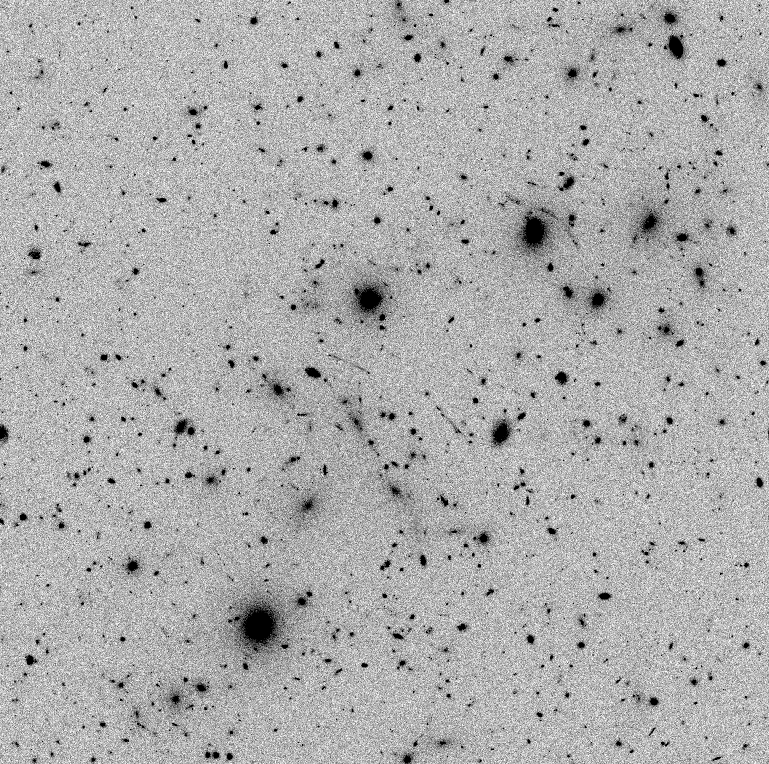}
\caption{Simulated observations of {\em Ares} with {\em Euclid} in the broad {\tt riz} band (VIS channel). In the left panel, a single lens plane is used, corresponding to the convergence map shown in the first panel of Fig.~\ref{fig:effconv}. In the right panel, we stack three lens planes. The resulting effective convergence is shown in the fourth panel of Fig.~\ref{fig:effconv}.}
\label{fig:euclid}
\end{figure*}

\subsection{\emph{Euclid}}
Finally, we simulate an observation of {\em Ares} also with the future {\em Euclid} space telescope \citep{2011arXiv1110.3193L}. {\em Euclid} is scheduled for launch in the early 2020s and will observe $15,000$ sq. degrees of the sky in four bands, namely a broad {\tt {riz}} band (VIS), and three near infrared bands ({\tt{Y}}, {\tt{J}}, {\tt{H}}). In the VIS channel the spatial resolution will be $0.1"$ and the observations will reach a magnitude limit of $24.5$ $m_{\rm AB}$ for extended sources at $S/N\sim10$, making {\em Euclid} a very promising instrument for strong lensing science. 

We use the set-up of a typical {\em Euclid} observation to illustrate the multi-plane functionality of \sky. For these simulations, we assume a PSF with a Moffat profile \citep{1969A&A.....3..455M} and FWHM of 0.18". We produce two images in the VIS channel. The first includes only one lens plane, as in the previous examples shown in Fig.~\ref{fig_color}. The second, includes the effects of the other two lens planes shown in  the second and third upper panels of Fig.~\ref{fig:effconv}, in addition to that  containing the mass distribution of {\em Ares}. The images are shown in Fig.~\ref{fig:euclid}. The effective mass distributions of the lenses in the two cases correspond to the first and to the last upper panels in  Fig.~\ref{fig:effconv}. Clearly, the addition of other lens planes impacts on the resulting appearance of the strong lensing features. Some of the arcs  are dimmed or broken into smaller arclets, while some other sources happen to be more strongly distorted. Overall, several lensed images in the right panel appear shifted compared to the corresponding ones in the left panel. 

\section{Discussion}
\label{sec:discussion}
%Discuss applications of the algorithm.
In this section we discuss possible applications of \sky. Simulations represent a fundamental tool in gravitational lensing since there are no sources in the sky whose intrinsic shapes, before lensing, are perfectly known. Simulations with known inputs allow to calibrate lensing measurement codes and to quantify the impact of random errors (such as noise). In addition, they help to improve the understanding and characterization of systematic and modeling errors. This is useful when testing mass reconstruction codes that aim to constraint the distribution of the lens light, its mass distribution, and the background sources. In particular, the knots of substructure that represent regions of star formation included in this version of \sky\ will also produce multiple images in strongly-lensed arcs (e.g., \ref{f5}) that can place to better constraints during the source image reconstruction process. These knots of star formation are also important to better understand the images of regions of star formation in the high-redshift universe that are magnified by the effect of strong lensing. 

\sky\ can be used to simulate gravitational arcs in the center of galaxy clusters and to improve the modeling of strong lensing systems, crucial to exploit the information that they contain about the distribution of dark matter in galaxies and galaxy clusters. This is also important to measure the shape of dark matter halos with precision and to detect and constrain their substructures. 

{The code can also be used to simulate lensing effects around galaxies, although this was not extensively discussed in this paper \citep{metcalf18}.} Future wide galaxy surveys such as the Large Synoptic Survey Telescope (LSST) \citep{ivezic08} and \emph{Euclid} will increase the number of strong lensing systems by up to three orders of magnitude compared to current searches (e.g., there will be about 120000 and 170000 galaxy-galaxy strong
lenses in the cases of LSST and \emph{Euclid}, respectively \citet{collet15}). \sky\ simulations can be used to develop and train algorithms to find these systems with the efficiency and completeness required. 

\section{Summary and Conclusions}
\label{sec:conclusions}

We have presented a new version of the ray-tracing code \sky. New additions to the code include the use of denoised source galaxies from the Hubble eXtreme Deep Field, which improves the resolution of the background source images as compared with previous versions of \sky. This new version of \sky\ is able to calculate the lensing effects due to multiple deflector planes along the line of sight, and also is able to simulate substructures in the source images that represent regions of star formation. We model these substructures as objects with a S\'ersic profile, and we perform a morphological analysis of images of nearby galaxies to empirically motivate their size and luminosity distributions.  
\sky\ is able to produce observations of the images of background sources lensed by multiple deflectors along the line-of-sight through virtually any instrument and telescope---both ground- and space-based. The mass distribution of the lenses (produced either analytically or numerically) is used to generate a deflection angle field per lens plane. {Multiple lens planes can be used to include lensing effects by any mass along the line of sight.} Once the total throughput of the system, a model for the PSF, and the characteristics of the instrument and telescope are specified, \sky\ will produce a simulated image of an observation (including noise) for a given field-of-view and exposure time.

As an example of the simulations that can be generated with \sky\, we have created images in multiple filters for several imaging instruments of current and future space and ground-based observatories: \emph{HST} ACS WFC, \emph{WFIRST} WFI, \emph{JWST} NIRcam, Subaru's HSC, and \emph{Euclid}'s VIS. Simulations  with known input allow for the testing, calibration, and improvement of lens modeling algorithms, in addition to providing a means to tests for the impact of different sources of random and systematic errors. The regions of star formation included in the source images of this version of \sky\ will allow, for example, to provide better constraints on source image reconstructions and lens models. The simulations created by \sky\ can also be utilized to train algorithms that automatize the search for strong lensing systems in large data sets. 

{We will make \sky\ available through the {\em Bologna Lens Factory} portal.\footnote{\url{http://metcalf1.difa.unibo.it/blf-portal/skylens.html}.}}

%used by lensing inversion algorithms, providing  additional constraints that limit the parameter space of solutions
%*summary of uses 
%*future work? 
%*distribution

\section*{Acknowledgements}
 We acknowledge support from the Italian Ministry of Foreign Affairs and International Cooperation (MAECI), Directorate General for Country Promotion, from ASI via contract ASI/INAF/I/023/12/0. We acknowledge support from grant HST-AR-13915.004-A of the Space Telescope Science Institute. AAP is supported by the Jet Propulsion Laboratory.  JR is being supported in part by the Jet Propulsion Laboratory. The research was carried out in part at the Jet Propulsion Laboratory, California Institute of Technology, under a contract with the National Aeronautics and Space Administration. M. Maturi was supported by the SFB-Transregio TR33 ``The Dark Universe".
\textcopyright 2017. All rights reserved.

\bibliographystyle{mnras}
\bibliography{skylens3_paper}

\begin{thebibliography}{}
\makeatletter
\relax
\def\mn@urlcharsother{\let\do\@makeother \do\$\do\&\do\#\do\^\do\_\do\%\do\~}
\def\mn@doi{\begingroup\mn@urlcharsother \@ifnextchar [ {\mn@doi@}
  {\mn@doi@[]}}
\def\mn@doi@[#1]#2{\def\@tempa{#1}\ifx\@tempa\@empty \href
  {http://dx.doi.org/#2} {doi:#2}\else \href {http://dx.doi.org/#2} {#1}\fi
  \endgroup}
\def\mn@eprint#1#2{\mn@eprint@#1:#2::\@nil}
\def\mn@eprint@arXiv#1{\href {http://arxiv.org/abs/#1} {{\tt arXiv:#1}}}
\def\mn@eprint@dblp#1{\href {http://dblp.uni-trier.de/rec/bibtex/#1.xml}
  {dblp:#1}}
\def\mn@eprint@#1:#2:#3:#4\@nil{\def\@tempa {#1}\def\@tempb {#2}\def\@tempc
  {#3}\ifx \@tempc \@empty \let \@tempc \@tempb \let \@tempb \@tempa \fi \ifx
  \@tempb \@empty \def\@tempb {arXiv}\fi \@ifundefined
  {mn@eprint@\@tempb}{\@tempb:\@tempc}{\expandafter \expandafter \csname
  mn@eprint@\@tempb\endcsname \expandafter{\@tempc}}}

\bibitem[\protect\citeauthoryear{{Aihara} et~al.,}{{Aihara}
  et~al.}{2017}]{aihara17}
{Aihara} H.,  et~al., 2017, preprint, \href
  {http://adsabs.harvard.edu/abs/2017arXiv170208449A} {} (\mn@eprint {arXiv}
  {1702.08449})

\bibitem[\protect\citeauthoryear{{Allen}, {Evrard}  \& {Mantz}}{{Allen}
  et~al.}{2011}]{allen11}
{Allen} S.~W.,  {Evrard} A.~E.,   {Mantz} A.~B.,  2011, \mn@doi [\araa]
  {10.1146/annurev-astro-081710-102514}, \href
  {http://adsabs.harvard.edu/abs/2011ARA%26A..49..409A} {49, 409}

\bibitem[\protect\citeauthoryear{{Amara}, {Metcalf}, {Cox}  \&
  {Ostriker}}{{Amara} et~al.}{2006}]{amara06}
{Amara} A.,  {Metcalf} R.~B.,  {Cox} T.~J.,   {Ostriker} J.~P.,  2006, \mn@doi
  [\mnras] {10.1111/j.1365-2966.2006.10053.x}, \href
  {http://adsabs.harvard.edu/abs/2006MNRAS.367.1367A} {367, 1367}

\bibitem[\protect\citeauthoryear{{Bailey}}{{Bailey}}{2012}]{bailey12}
{Bailey} S.,  2012, \mn@doi [\pasp] {10.1086/668105}, \href
  {http://adsabs.harvard.edu/abs/2012PASP..124.1015B} {124, 1015}

\bibitem[\protect\citeauthoryear{{Bartelmann} \& {Maturi}}{{Bartelmann} \&
  {Maturi}}{2017}]{bartelmann17}
{Bartelmann} M.,  {Maturi} M.,  2017, \mn@doi [Scholarpedia]
  {10.4249/scholarpedia.32440}, \href
  {http://adsabs.harvard.edu/abs/2017SchpJ..1232440B} {12, 32440}

\bibitem[\protect\citeauthoryear{{Bartelmann} \& {Schneider}}{{Bartelmann} \&
  {Schneider}}{2001}]{bartelmann01}
{Bartelmann} M.,  {Schneider} P.,  2001, \mn@doi [\physrep]
  {10.1016/S0370-1573(00)00082-X}, \href
  {http://adsabs.harvard.edu/abs/2001PhR...340..291B} {340, 291}

\bibitem[\protect\citeauthoryear{{Beckwith} et~al.,}{{Beckwith}
  et~al.}{2006}]{beckwith06}
{Beckwith} S.~V.~W.,  et~al., 2006, \mn@doi [\aj] {10.1086/507302}, \href
  {http://adsabs.harvard.edu/abs/2006AJ....132.1729B} {132, 1729}

\bibitem[\protect\citeauthoryear{{Ben{\'{\i}}tez}}{{Ben{\'{\i}}tez}}{2000}]{benitez00}
{Ben{\'{\i}}tez} N.,  2000, \mn@doi [\apj] {10.1086/308947}, \href
  {http://adsabs.harvard.edu/abs/2000ApJ...536..571B} {536, 571}

\bibitem[\protect\citeauthoryear{{Ben{\'{\i}}tez} et~al.,}{{Ben{\'{\i}}tez}
  et~al.}{2004}]{benitez04}
{Ben{\'{\i}}tez} N.,  et~al., 2004, \mn@doi [\apjs] {10.1086/380120}, \href
  {http://adsabs.harvard.edu/abs/2004ApJS..150....1B} {150, 1}

\bibitem[\protect\citeauthoryear{{Bernstein} \& {Jarvis}}{{Bernstein} \&
  {Jarvis}}{2002}]{bernstein02}
{Bernstein} G.~M.,  {Jarvis} M.,  2002, \mn@doi [\aj] {10.1086/338085}, \href
  {http://adsabs.harvard.edu/abs/2002AJ....123..583B} {123, 583}

\bibitem[\protect\citeauthoryear{{Bertin}}{{Bertin}}{2006}]{bertin06}
{Bertin} E.,  2006, in {Gabriel} C.,  {Arviset} C.,  {Ponz} D.,   {Enrique} S.,
   eds,  Astronomical Society of the Pacific Conference Series Vol. 351,
  Astronomical Data Analysis Software and Systems XV. p.~112

\bibitem[\protect\citeauthoryear{{Birrer} \& {Amara}}{{Birrer} \&
  {Amara}}{2018}]{birrer18}
{Birrer} S.,  {Amara} A.,  2018, preprint, \href
  {http://adsabs.harvard.edu/abs/2018arXiv180309746B} {} (\mn@eprint {arXiv}
  {1803.09746})

\bibitem[\protect\citeauthoryear{{Caminha} et~al.,}{{Caminha}
  et~al.}{2017}]{caminha17}
{Caminha} G.~B.,  et~al., 2017, \mn@doi [\aap] {10.1051/0004-6361/201629297},
  \href {http://adsabs.harvard.edu/abs/2017A%26A...600A..90C} {600, A90}

\bibitem[\protect\citeauthoryear{{Carrasco}, {Stapelberg}, {Maturi},
  {Bartelmann}, {Seidel}  \& {Erben}}{{Carrasco} et~al.}{2018}]{carrasco18}
{Carrasco} M.,  {Stapelberg} S.,  {Maturi} M.,  {Bartelmann} M.,  {Seidel} G.,
   {Erben} T.,  2018, preprint, \href
  {http://adsabs.harvard.edu/abs/2018arXiv180703793C} {} (\mn@eprint {arXiv}
  {1807.03793})

\bibitem[\protect\citeauthoryear{{Coe}, {Ben{\'{\i}}tez}, {S{\'a}nchez}, {Jee},
  {Bouwens}  \& {Ford}}{{Coe} et~al.}{2006}]{coe06}
{Coe} D.,  {Ben{\'{\i}}tez} N.,  {S{\'a}nchez} S.~F.,  {Jee} M.,  {Bouwens} R.,
    {Ford} H.,  2006, \mn@doi [\aj] {10.1086/505530}, \href
  {http://adsabs.harvard.edu/abs/2006AJ....132..926C} {132, 926}

\bibitem[\protect\citeauthoryear{{Coe} et~al.,}{{Coe} et~al.}{2012}]{coe12}
{Coe} D.,  et~al., 2012, \mn@doi [\apj] {10.1088/0004-637X/757/1/22}, \href
  {http://adsabs.harvard.edu/abs/2012ApJ...757...22C} {757, 22}

\bibitem[\protect\citeauthoryear{{Collett}}{{Collett}}{2015}]{collet15}
{Collett} T.~E.,  2015, \mn@doi [\apj] {10.1088/0004-637X/811/1/20}, \href
  {http://adsabs.harvard.edu/abs/2015ApJ...811...20C} {811, 20}

\bibitem[\protect\citeauthoryear{{Frei}}{{Frei}}{1996}]{frei96b}
{Frei} Z.,  1996, \mn@doi [\pasp] {10.1086/133775}, \href
  {http://adsabs.harvard.edu/abs/1996PASP..108..624F} {108, 624}

\bibitem[\protect\citeauthoryear{{Frei}, {Guhathakurta}, {Gunn}  \&
  {Tyson}}{{Frei} et~al.}{1996}]{frei96}
{Frei} Z.,  {Guhathakurta} P.,  {Gunn} J.~E.,   {Tyson} J.~A.,  1996, \mn@doi
  [\aj] {10.1086/117771}, \href
  {http://adsabs.harvard.edu/abs/1996AJ....111..174F} {111, 174}

\bibitem[\protect\citeauthoryear{{Giocoli}, {Meneghetti}, {Bartelmann},
  {Moscardini}  \& {Boldrin}}{{Giocoli} et~al.}{2012}]{giocoli12}
{Giocoli} C.,  {Meneghetti} M.,  {Bartelmann} M.,  {Moscardini} L.,   {Boldrin}
  M.,  2012, \mn@doi [\mnras] {10.1111/j.1365-2966.2012.20558.x}, \href
  {http://adsabs.harvard.edu/abs/2012MNRAS.421.3343G} {421, 3343}

\bibitem[\protect\citeauthoryear{{Grazian}, {Fontana}, {De Santis}, {Gallozzi},
  {Giallongo}  \& {Di Pangrazio}}{{Grazian} et~al.}{2004}]{grazian04}
{Grazian} A.,  {Fontana} A.,  {De Santis} C.,  {Gallozzi} S.,  {Giallongo} E.,
   {Di Pangrazio} F.,  2004, \mn@doi [\pasp] {10.1086/423123}, \href
  {http://adsabs.harvard.edu/abs/2004PASP..116..750G} {116, 750}

\bibitem[\protect\citeauthoryear{{Heymans} et~al.,}{{Heymans}
  et~al.}{2006}]{heymans06}
{Heymans} C.,  et~al., 2006, \mn@doi [\mnras]
  {10.1111/j.1365-2966.2006.10198.x}, \href
  {http://adsabs.harvard.edu/abs/2006MNRAS.368.1323H} {368, 1323}

\bibitem[\protect\citeauthoryear{{Hirata}, {Gehrels}, {Kneib}, {Kruk},
  {Rhodes}, {Wang}  \& {Zoubian}}{{Hirata} et~al.}{2012}]{hirata12}
{Hirata} C.~M.,  {Gehrels} N.,  {Kneib} J.-P.,  {Kruk} J.,  {Rhodes} J.,
  {Wang} Y.,   {Zoubian} J.,  2012, preprint, \href
  {http://adsabs.harvard.edu/abs/2012arXiv1204.5151H} {} (\mn@eprint {arXiv}
  {1204.5151})

\bibitem[\protect\citeauthoryear{{Huang} et~al.,}{{Huang}
  et~al.}{2016}]{huang16}
{Huang} K.-H.,  et~al., 2016, \mn@doi [\apjl] {10.3847/2041-8205/823/1/L14},
  \href {http://adsabs.harvard.edu/abs/2016ApJ...823L..14H} {823, L14}

\bibitem[\protect\citeauthoryear{{Illingworth} et~al.,}{{Illingworth}
  et~al.}{2013}]{illingworth13}
{Illingworth} G.~D.,  et~al., 2013, \mn@doi [\apjs]
  {10.1088/0067-0049/209/1/6}, \href
  {http://adsabs.harvard.edu/abs/2013ApJS..209....6I} {209, 6}

\bibitem[\protect\citeauthoryear{{Ivezic} et~al.,}{{Ivezic}
  et~al.}{2008}]{ivezic08}
{Ivezic} Z.,  et~al., 2008, preprint, \href
  {http://adsabs.harvard.edu/abs/2008arXiv0805.2366I} {} (\mn@eprint {arXiv}
  {0805.2366})

\bibitem[\protect\citeauthoryear{{Johnson} et~al.,}{{Johnson}
  et~al.}{2017a}]{johnson17a}
{Johnson} T.~L.,  et~al., 2017a, \mn@doi [\apj] {10.3847/1538-4357/aa7756},
  \href {http://adsabs.harvard.edu/abs/2017ApJ...843...78J} {843, 78}

\bibitem[\protect\citeauthoryear{{Johnson} et~al.,}{{Johnson}
  et~al.}{2017b}]{johnson17b}
{Johnson} T.~L.,  et~al., 2017b, \mn@doi [\apjl] {10.3847/2041-8213/aa7516},
  \href {http://adsabs.harvard.edu/abs/2017ApJ...843L..21J} {843, L21}

\bibitem[\protect\citeauthoryear{{Kannawadi}, {Shapiro}, {Mandelbaum},
  {Hirata}, {Kruk}  \& {Rhodes}}{{Kannawadi} et~al.}{2015}]{kannawadi15}
{Kannawadi} A.,  {Shapiro} C.~A.,  {Mandelbaum} R.,  {Hirata} C.~M.,  {Kruk}
  J.~W.,   {Rhodes} J.~D.,  2015, preprint, \href
  {http://adsabs.harvard.edu/abs/2015arXiv151201570K} {} (\mn@eprint {arXiv}
  {1512.01570})

\bibitem[\protect\citeauthoryear{{Kelly} et~al.,}{{Kelly}
  et~al.}{2017}]{kelly17}
{Kelly} P.~L.,  et~al., 2017, preprint, \href
  {http://adsabs.harvard.edu/abs/2017arXiv170610279K} {} (\mn@eprint {arXiv}
  {1706.10279})

\bibitem[\protect\citeauthoryear{{Kennicutt}, {Edgar}  \& {Hodge}}{{Kennicutt}
  et~al.}{1989}]{kennicutt89}
{Kennicutt} Jr. R.~C.,  {Edgar} B.~K.,   {Hodge} P.~W.,  1989, \mn@doi [\apj]
  {10.1086/167147}, \href {http://adsabs.harvard.edu/abs/1989ApJ...337..761K}
  {337, 761}

\bibitem[\protect\citeauthoryear{{Kilbinger}}{{Kilbinger}}{2015}]{kilbinger15}
{Kilbinger} M.,  2015, \mn@doi [Reports on Progress in Physics]
  {10.1088/0034-4885/78/8/086901}, \href
  {http://adsabs.harvard.edu/abs/2015RPPh...78h6901K} {78, 086901}

\bibitem[\protect\citeauthoryear{{Kitching} et~al.,}{{Kitching}
  et~al.}{2013}]{kitching13}
{Kitching} T.~D.,  et~al., 2013, \mn@doi [\apjs] {10.1088/0067-0049/205/2/12},
  \href {http://adsabs.harvard.edu/abs/2013ApJS..205...12K} {205, 12}

\bibitem[\protect\citeauthoryear{{Krist}, {Hook}  \& {Stoehr}}{{Krist}
  et~al.}{2011}]{krist11}
{Krist} J.~E.,  {Hook} R.~N.,   {Stoehr} F.,  2011, in Optical Modeling and
  Performance Predictions V. p. 81270J, \mn@doi{10.1117/12.892762}

\bibitem[\protect\citeauthoryear{{Laureijs} et~al.,}{{Laureijs}
  et~al.}{2011}]{2011arXiv1110.3193L}
{Laureijs} R.,  et~al., 2011, preprint, \href
  {http://adsabs.harvard.edu/abs/2011arXiv1110.3193L} {} (\mn@eprint {arXiv}
  {1110.3193})

\bibitem[\protect\citeauthoryear{{Li}, {Gladders}, {Rangel}, {Florian},
  {Bleem}, {Heitmann}, {Habib}  \& {Fasel}}{{Li} et~al.}{2016}]{li16}
{Li} N.,  {Gladders} M.~D.,  {Rangel} E.~M.,  {Florian} M.~K.,  {Bleem} L.~E.,
  {Heitmann} K.,  {Habib} S.,   {Fasel} P.,  2016, \mn@doi [\apj]
  {10.3847/0004-637X/828/1/54}, \href
  {http://adsabs.harvard.edu/abs/2016ApJ...828...54L} {828, 54}

\bibitem[\protect\citeauthoryear{{Livermore}, {Finkelstein}  \&
  {Lotz}}{{Livermore} et~al.}{2017}]{livermore17}
{Livermore} R.~C.,  {Finkelstein} S.~L.,   {Lotz} J.~M.,  2017, \mn@doi [\apj]
  {10.3847/1538-4357/835/2/113}, \href
  {http://adsabs.harvard.edu/abs/2017ApJ...835..113L} {835, 113}

\bibitem[\protect\citeauthoryear{{Lotz}}{{Lotz}}{2015}]{lotz15}
{Lotz} J.,  2015, IAU General Assembly, \href
  {http://adsabs.harvard.edu/abs/2015IAUGA..2255460L} {22, 2255460}

\bibitem[\protect\citeauthoryear{{Maturi}}{{Maturi}}{2016}]{maturi16}
{Maturi} M.,  2016, preprint, \href
  {http://adsabs.harvard.edu/abs/2016arXiv160705724M} {} (\mn@eprint {arXiv}
  {1607.05724})

\bibitem[\protect\citeauthoryear{{Meneghetti} et~al.,}{{Meneghetti}
  et~al.}{2008}]{meneghetti08}
{Meneghetti} M.,  et~al., 2008, \mn@doi [\aap] {10.1051/0004-6361:20079119},
  \href {http://adsabs.harvard.edu/abs/2008A%26A...482..403M} {482, 403}

\bibitem[\protect\citeauthoryear{{Meneghetti}, {Rasia}, {Merten}, {Bellagamba},
  {Ettori}, {Mazzotta}, {Dolag}  \& {Marri}}{{Meneghetti}
  et~al.}{2010}]{meneghetti10a}
{Meneghetti} M.,  {Rasia} E.,  {Merten} J.,  {Bellagamba} F.,  {Ettori} S.,
  {Mazzotta} P.,  {Dolag} K.,   {Marri} S.,  2010, \mn@doi [\aap]
  {10.1051/0004-6361/200913222}, \href
  {http://adsabs.harvard.edu/abs/2010A%26A...514A..93M} {514, A93}

\bibitem[\protect\citeauthoryear{{Meneghetti} et~al.,}{{Meneghetti}
  et~al.}{2017}]{meneghetti17}
{Meneghetti} M.,  et~al., 2017, \mn@doi [\mnras] {10.1093/mnras/stx2064}, \href
  {http://adsabs.harvard.edu/abs/2017MNRAS.472.3177M} {472, 3177}

\bibitem[\protect\citeauthoryear{{Metcalf} et~al.,}{{Metcalf}
  et~al.}{2018}]{metcalf18}
{Metcalf} R.~B.,  et~al., 2018, preprint, \href
  {http://adsabs.harvard.edu/abs/2018arXiv180203609M} {} (\mn@eprint {arXiv}
  {1802.03609})

\bibitem[\protect\citeauthoryear{{Moffat}}{{Moffat}}{1969}]{1969A&A.....3..455M}
{Moffat} A.~F.~J.,  1969, \aap, \href
  {http://adsabs.harvard.edu/abs/1969A%26A.....3..455M} {3, 455}

\bibitem[\protect\citeauthoryear{{Petkova}, {Metcalf}  \& {Giocoli}}{{Petkova}
  et~al.}{2014}]{2014MNRAS.445.1954P}
{Petkova} M.,  {Metcalf} R.~B.,   {Giocoli} C.,  2014, \mn@doi [\mnras]
  {10.1093/mnras/stu1860}, \href
  {http://adsabs.harvard.edu/abs/2014MNRAS.445.1954P} {445, 1954}

\bibitem[\protect\citeauthoryear{{Rafelski} et~al.,}{{Rafelski}
  et~al.}{2015}]{rafelski15}
{Rafelski} M.,  et~al., 2015, \mn@doi [\aj] {10.1088/0004-6256/150/1/31}, \href
  {http://adsabs.harvard.edu/abs/2015AJ....150...31R} {150, 31}

\bibitem[\protect\citeauthoryear{{Rasia} et~al.,}{{Rasia}
  et~al.}{2012}]{rasia12}
{Rasia} E.,  et~al., 2012, \mn@doi [New Journal of Physics]
  {10.1088/1367-2630/14/5/055018}, \href
  {http://adsabs.harvard.edu/abs/2012NJPh...14e5018R} {14, 055018}

\bibitem[\protect\citeauthoryear{{Refregier}}{{Refregier}}{2003}]{refregier03b}
{Refregier} A.,  2003, \mn@doi [\mnras] {10.1046/j.1365-8711.2003.05901.x},
  \href {http://adsabs.harvard.edu/abs/2003MNRAS.338...35R} {338, 35}

\bibitem[\protect\citeauthoryear{{Rowe} et~al.,}{{Rowe} et~al.}{2015}]{rowe15}
{Rowe} B.~T.~P.,  et~al., 2015, \mn@doi [Astronomy and Computing]
  {10.1016/j.ascom.2015.02.002}, \href
  {http://adsabs.harvard.edu/abs/2015A%26C....10..121R} {10, 121}

\bibitem[\protect\citeauthoryear{{Schneider}, {Gunn}  \& {Hoessel}}{{Schneider}
  et~al.}{1983}]{schneider83}
{Schneider} D.~P.,  {Gunn} J.~E.,   {Hoessel} J.~G.,  1983, \mn@doi [\apj]
  {10.1086/160602}, \href {http://adsabs.harvard.edu/abs/1983ApJ...264..337S}
  {264, 337}

\bibitem[\protect\citeauthoryear{{Seidel} \& {Bartelmann}}{{Seidel} \&
  {Bartelmann}}{2007}]{seidel07}
{Seidel} G.,  {Bartelmann} M.,  2007, \mn@doi [\aap]
  {10.1051/0004-6361:20066097}, \href
  {http://adsabs.harvard.edu/abs/2007A%26A...472..341S} {472, 341}

\bibitem[\protect\citeauthoryear{{S{\'e}rsic}}{{S{\'e}rsic}}{1963}]{sersic63}
{S{\'e}rsic} J.~L.,  1963, Boletin de la Asociacion Argentina de Astronomia La
  Plata Argentina, \href {http://adsabs.harvard.edu/abs/1963BAAA....6...41S}
  {6, 41}

\bibitem[\protect\citeauthoryear{{Sokol}, {Anderson}  \& {Smith}}{{Sokol}
  et~al.}{2012}]{sokol12}
{Sokol} J.,  {Anderson} J.,   {Smith} L.,  2012, Technical report, {Assessing
  ACS/WFC Sky Backgrounds}

\bibitem[\protect\citeauthoryear{{Stapelberg}, {Carrasco}  \&
  {Maturi}}{{Stapelberg} et~al.}{2017}]{stapelberg17}
{Stapelberg} S.,  {Carrasco} M.,   {Maturi} M.,  2017, preprint, \href
  {http://adsabs.harvard.edu/abs/2017arXiv170909758S} {} (\mn@eprint {arXiv}
  {1709.09758})

\bibitem[\protect\citeauthoryear{{Tessore}, {Bellagamba}  \&
  {Metcalf}}{{Tessore} et~al.}{2016}]{tessore16}
{Tessore} N.,  {Bellagamba} F.,   {Metcalf} R.~B.,  2016, \mn@doi [\mnras]
  {10.1093/mnras/stw2212}, \href
  {http://adsabs.harvard.edu/abs/2016MNRAS.463.3115T} {463, 3115}

\bibitem[\protect\citeauthoryear{{Thuan} \& {Gunn}}{{Thuan} \&
  {Gunn}}{1976}]{thuan76}
{Thuan} T.~X.,  {Gunn} J.~E.,  1976, \mn@doi [\pasp] {10.1086/129982}, \href
  {http://adsabs.harvard.edu/abs/1976PASP...88..543T} {88, 543}

\bibitem[\protect\citeauthoryear{{Treu}}{{Treu}}{2010}]{treu10}
{Treu} T.,  2010, \mn@doi [\araa] {10.1146/annurev-astro-081309-130924}, \href
  {http://adsabs.harvard.edu/abs/2010ARA%26A..48...87T} {48, 87}

\bibitem[\protect\citeauthoryear{{Vanzella} et~al.,}{{Vanzella}
  et~al.}{2017}]{vanzella17}
{Vanzella} E.,  et~al., 2017, \mn@doi [\mnras] {10.1093/mnras/stx351}, \href
  {http://adsabs.harvard.edu/abs/2017MNRAS.467.4304V} {467, 4304}

\bibitem[\protect\citeauthoryear{{Wade}, {Hoessel}, {Elias}  \&
  {Huchra}}{{Wade} et~al.}{1979}]{wade79}
{Wade} R.~A.,  {Hoessel} J.~G.,  {Elias} J.~H.,   {Huchra} J.~P.,  1979,
  \mn@doi [\pasp] {10.1086/130435}, \href
  {http://adsabs.harvard.edu/abs/1979PASP...91...35W} {91, 35}

\bibitem[\protect\citeauthoryear{{Weinberg}, {Mortonson}, {Eisenstein},
  {Hirata}, {Riess}  \& {Rozo}}{{Weinberg} et~al.}{2013}]{weinberg13}
{Weinberg} D.~H.,  {Mortonson} M.~J.,  {Eisenstein} D.~J.,  {Hirata} C.,
  {Riess} A.~G.,   {Rozo} E.,  2013, \mn@doi [\physrep]
  {10.1016/j.physrep.2013.05.001}, \href
  {http://adsabs.harvard.edu/abs/2013PhR...530...87W} {530, 87}

\bibitem[\protect\citeauthoryear{{de Vaucouleurs}, {de Vaucouleurs}, {Corwin},
  {Buta}, {Paturel}  \& {Fouqu{\'e}}}{{de Vaucouleurs}
  et~al.}{1991}]{devaucouleurs91}
{de Vaucouleurs} G.,  {de Vaucouleurs} A.,  {Corwin} Jr. H.~G.,  {Buta} R.~J.,
  {Paturel} G.,   {Fouqu{\'e}} P.,  1991, {Third Reference Catalogue of Bright
  Galaxies. Volume I: Explanations and references. Volume II: Data for galaxies
  between 0$^{h}$ and 12$^{h}$. Volume III: Data for galaxies between 12$^{h}$
  and 24$^{h}$.}

\makeatother
\end{thebibliography}
\end{document}